# Optical constants of DC sputtering derived ITO, TiO$_2$ and TiO$_2$:Nb thin films characterized by spectrophotometry and spectroscopic ellipsometry for optoelectronic devices

[a]Mohammed RASHEED, [b]RÉGIS BARILLÉ
[a,b]MOLTECH-Anjou, Universitéd'Angers/UMR CNRS 6200, 2 Bd Lavoisier, 49045 Angers, France
corresponding email: rasheed.mohammed40@yahoo.com

**Abstract**

Thin films of inorganic materials as Tin-doped indium oxide, titanium oxide, Niobium doped titanium oxide, were deposited for comparison on glass and Polyethylene terephthalate (PET) substrates with a DC sputtering method. These thin films have been characterized by different techniques: Dektak Surface Profilometer, X-ray diffraction (XRD), SEM, (UV/Vis/NIR) spectrophotometer and spectroscopic ellipsometry (SE). The optical parameters of these films such as transmittance, reflectance, refractive index, extinction coefficient, energy gap obtained with different electronic transitions, real and imaginary ($\varepsilon_r, \varepsilon_i$) dielectric constants, were determined in the wavelengths range of (200 - 2200) nm. The results were compared with SE measurements in the ranges of (0.56- 6.19) eV by a new amorphous model with steps of 1 nm. SE measurements of optical constant have been examined and confirm the accuracy of the (UV/Vis/NIR) results. The optical properties indicate an excellent transmittance in the visible range of (400 - 800) nm. The average transmittance of films on glass is about (86%, 91%, 85%) for (ITO, TiO$_2$, TiO$_2$:Nb (NTO)) respectively and decreases to about (85%, 81%, 82%) for PET substrates. For all these materials the optical band gap for direct transition was (3.53, 3.3, 3.6) eV on glass substrates and on PET substrates using two methods (UV and SE). A comparison between optical constants and thickness of these ultrathin films observed gives an excellent agreement with the UV results. The deposited films were also analyzed by XRD and showed an amorphous structure. The structural morphology of these thin films has been investigated and compared.

## Introduction

ITO, TiO$_2$ and TiO$_2$:Nb (NTO) have been used as transparent electrodes in many optoelectronic devices such as: Dye synthesized solar cells (DSSC) [1, 2, 3], organic emitting diode and devices (OLEDs) [4, 5, 6], liquid crystal display (LCDs) [7, 8]. In addition, these films have wide range of well perspective applications using flexible substrates such as: optoelectronic devices [9, 10], flat panel displays [11, 12, 13], photovoltaic devices; DSSC [14, 15, 16, 17, 18, 19], organic solar cells [20, 21, 22], organic light emitting diode [23, 24, 25], gas sensors [26, 27], light emitting transistor [28], photocatalytic [29, 30], electrochromic [31, 32], perovskite solar cells [33, 34], microelectronic devices [35], and thin film transistor [36, 37]. Transparent thin films are usually fabricated by using different techniques, such as: sol-gel (spin coating) [38, 39], spray pyrolysis [40, 41, 42], reactive magnetron sputtering [43, 44, 45], chemical vapor deposition (LPCVD) [46, 47, 48], pulse laser deposition (PLD) [4, 49, 50], and RF magnetron sputtering [51, 52] etc. Researchers have an increasing interest in using DC sputtering technique to prepare transparent thin films with lower sputter voltage and obtain thin film at room temperature, in this way there is no increasing of substrate temperature. Flexible organic Polyethylene terephthalate (PET) has been chosen in this study as substrate because it has many advantages such as low cost, temperature stability, chemical and moisture resistance and durability. PET has a lowest price compare with other flexible plastic substrates as polyimide and glass. It matches deposition conditions for thin films solar cells (temperature, degassing, etc.) and the melting temperature degree of PET of about 250 – 260 °C. Nevertheless, most of the thin films deposited on plastic substrate (PET) at room temperature deposition such as dc sputtering technique to prevent the PET substrates from damage due to the poor thermal ability of plastic substrates. PET is unsupported and a semi-crystalline thermo-plastic polyester derived from polyethylene terephthalate. Its excellent wear resistance, low coefficient of friction, high flexural modulus, and superior dimensional stability make it a versatile material for designing mechanical and electro-mechanical parts. The deposition conditions such as sputter power and pressure played an important role in film properties. The high quality of films on PET substrates gives the possibility as alternative substrates to replace the standard glasses. The electrical and optical properties of target-substrate are strongly depended on the deposition condition rate. However, a very accurate control of process parameters is necessary [53]. Choosing the PET films depends on the balance properties between a good flexibility, flatness, dimensional stability, high mechanical strength and cost. Three transparent thin films were prepared in this paper, ITO,

TiO$_2$ and Niobium-doped TiO$_2$ (NTO) all of them deposited by DC sputtering on both glass and PET substrates. This method includes the deposition of oxide films in open atmosphere. There are many studies on ITO, TiO$_2$ and (NTO) Thin films obtained by DC sputtering technique [54, 55, 56]. This is why it is important to review and give a complete optical characterization of these materials.

Up to now, many researchers have investigated the optical constants of these films by spectroscopic ellipsometry (SE) and spectrophotometry (UV) for films deposited on glass and PET substrates. For example, **Tsai Shu-Yi et al.** [57] studied and compared the optical properties of Zinc Oxide films deposited by reactive RF magnetron sputtering on glass and PET substrates. The average transmittance of these films was above 80% in the range of 350-850 nm for each ZnO film deposited on glass or PET substrates. The value of the optical energy gap for 460 nm film thickness has the same value for each ZnO film deposited on different substrates. **Demiryont H. et al.** [58] characterized the optical properties of TiO$_2$ in particular the transmission spectrum and the refractive index of films with a thickness of 200-400 nm by ellipsometry and spectrophotometry within the range of 0.4 – 2.4 µm. The thin films were deposited by ion-beam sputtering of a metallic target on silicon substrates. **Minami T. et al.** [59] demonstrated a very thin ITO film preparation deposited onto glass and PET substrates by DC sputtering with a low temperature process in pure argon gas. They found a high transmittance of the films of about 97% in the visible region. **Kulkarni A. K. et al.** [60] published a paper in which they described the optical transmittance of ITO thin films with a thickness of about 50-100 nm by rf sputter deposited on glass and PET substrates. As a result, they found that the optical transmittance is about 60 - 70% in the visible range for the studied samples. **Faraj M. G. et al.** [61] reported the structural, thermal and optical properties of PET using different experimental methods. The average transmittance of the PET substrate was about 85% in the visible region making PET substrates suitable and useful for flexible optoelectronic applications and giving them the possibility to be used as alternative substrates in the goal to replace conventional glasses. **Fanta et al**. [62] reported the optical characterization of TiO$_2$ thin films prepared by magnetron sputtering onto K64 glass substrate within the wavelength range of 230 - 1000 nm by both spectroscopic photometry and spectroscopic ellipsometry using Caushy and Urbach formulas. They found that the refractive index of TiO$_2$ films are homogeneous and have an uniform thickness.

We currently did not find any reports for ITO, TiO$_2$ and NTO films fabricated by DC sputtering and deposited on glass and PET substrates giving the optical parameters of these thin films and studied using two different methods (UV and SE) with a comparison between them. In this goal in this paper the morphology, structure, growth and optical constants of deposited films **of** ITO, TiO$_2$, NTO thin films deposited on glass and PET substrates at room temperature with a DC sputtering technique were investigated and compared using spectrophotometry (UV) and spectroscopic ellipsometry (SE). The objective of the present work is to investigate the optical constants of the thin films using two different techniques of characterization (UV and SE). These results obtained for thin films deposited on different substrates (glass and PET) are useful and very important in order to check and confirm the variation of optical constants in the visible region. A wide band gap is the cause of a high transparency in the visible region of these thin films, making these films excellent and promising for transparent window materials such as solar cell applications.

**Experimental Details**

The DC sputtering technique is used to deposit ITO, TiO$_2$ and NTO transparent thin films on both glass and PET substrates with identical conditions. ITO ultrathin films were deposited in reactive atmosphere using In$_2$O$_3$:SnO$_2$ composition of $In_2O_3:SnO_2$ composition of $90:10\ wt\%$ respectively. Commercial plane-glass microscope slides were cut into $25x20$ mm plates and used as substrates. Glass substrates were cleaned carefully in an ultrasonic bath treatment type (BRANSON Ultrasonic-CAMDA 19 spc) with ethanol, acetone, deionized water and dichloromethane each of them during 20 minutes and dried with a nitrogen gas jet. The glass and PET substrates were placed on a rotating disk in a vertical direction with respect to the target and kept at ambient temperature. The distance between the target and the substrate was kept at about $7\ cm$, the deposition rate and the vacuum pressure of these films was: $4\ nm/min$ and 5 Pa respectively. For $ITO$ films (the current I = 20 mA) while I = 100 mA for TiO$_2$ and NTO films with the sputtering power of 100W at room temperature. The thicknesses of these thin films were measured by Surface Profilometer with a Veeco 6$^M$ Metrology L.L.C and was 20.5±0.3 nm. The spectroscopic ellipsometry thickness range was 20.5±0.4 nm for each of these thin films. The structures of the samples were analyzed by D8 Advance Brucker diffractometer $CuK\alpha$ 1,2 ($\lambda = 1.5406\ A^o$) - (XRD) X-ray diffraction consisted of a linear Vantec super speed detector. The morphology of these thin films appeared amorphous and were analyzed using a SEM (Scanning electronic microscope: JEOL

Microscope). The optical properties: transmittance, reflectance, and absorption spectra were obtained at room temperature in the wavelength range of 200 - 2200 nm with a Perkin Elmer Lambda 950 (UV/Vis/NIR) Spectrophotometer. In addition, the refractive and extinction coefficients $(n, k)$, absorption coefficient $(\alpha)$, the band gap $(E_g)$, and real and imaginary dielectric constants $\varepsilon_r$ and $\varepsilon_i$ respectively were calculated. The Transmittance and reflectance were made in the same spectral range of the spectrophotometer using a phase modulated spectroscopic ellipsometry type (UV-Vis-NIR Horiba Jobin-Yvon); at $70^0$ angle of incident.

The spectroscopic ellipsometry (SE) measurements expressed by the complex reflectance ratio (ρ) as a function of wavelength which is described by the ratio between the perpendicular $(s -)$ and parallel $(p -)$ polarized light reflection coefficients called (Fresnel reflection coefficients) expressing the electrical field alteration after reflection [63, 64, 65, 66, 67, 68]

$$\rho = \frac{r_p}{r_s} = tan\psi e^{i\Delta} = tan\psi(cos\Delta + isin\Delta) = f(n, k, t) \qquad (1)$$

where $r_p$ and $r_s$ ratio of complex Fresnel reflection coefficients of the sample, for parallel $(p -)$ and perpendicular $(s -)$ polarized light with respect to the sample plane of incidence respectively. $\Delta = \delta_p - \delta_s$ is the phase shift difference, and $f(n, k, t)$ is an optical function dependent on the layer thickness.

In this study, the new amorphous model is defined by the following equations [37-39]-:

$$k(\omega) = \begin{cases} \frac{f_j.(\omega-\omega_g)^2}{(\omega-\omega_j)^2+\Gamma_j^2} & ; for\ \omega > \omega_g \\ 0 & ; for\ \omega \leq \omega_g \end{cases} \text{ and } n(\omega) = \begin{cases} n_\infty + \frac{B_j.(\omega-\omega_j)^2+c}{(\omega-\omega_j)^2-\Gamma_j^2} & ; for\ \omega > \omega_g \\ 0 & ; for\ \omega \leq \omega_g \end{cases}$$

where: $k(\omega)$ is the extinction coefficient, $n(\omega)$ is the refractive index, $B_j = \frac{f_j}{\Gamma_j}\left(\Gamma_j^2 - (\omega_j - \omega_g)^2\right)$ and $c_j = 2.f_j.\Gamma_j.(\omega - \omega_g)$, $n_\infty$ is the refractive index and equal to the value of the refractive index when $(\omega \to \infty)$, $f_j(eV)$ is the amplitude of the peak of the $k(\omega)$. $\Gamma_j\ (eV)$ is the broadening term of the peak of absorption. $\omega_j(eV)$ is the energy at which $k(\omega)$ is maximum. $\omega_g(eV)$ is the optical band gap.

## 3. Results and discussions

3.1 Structural and morphological analysis

The crystal structural of the thin films deposited on glass and PET substrates is examined by X-ray diffraction using a standard $Cuk\alpha$ radiation ($\lambda = 0.15406\ nm$), and are amorphous. The structural morphology of these thin films deposited on glass and PET substrates has been analyzed by scanning electron microscope (SEM). As we have observed the films on glass substrate have a surface smother than those on PET substrate. The larger grains of the films are absorbed on the PET substrates. For ITO thin films one can not recognize the difference between the crystallite growth of the ITO grains deposited on glass and PET substrates [see figure 1]. Obviously the influence of the substrate nature on the crystallite growth of the thin TiO$_2$ and NTO films deposited on both glass and PET substrates is clearly noticed, while the genesis of crystalline grain size on the PET gives larger grains compared to the same film deposited on glass substrates [see figure 1].

### 3.2 Optical characterization of ITO, TiO$_2$ and TiO$_2$:Nb (NTO) thin films

### 3.2.1(UV/vis/NIR) Spectrophotometry analysis

The transmission and reflection spectra were recorded between 200 nm and 2200 nm. The figure 2 gives the optical transmission spectra of ITO, TiO$_2$ and TiO$_2$:Nb (NTO) thin films deposited on glass (Fig. 3a) and PET (Fig. 3b) substrates. The visible average transmittance $T_{ave}$ is (86%, 91%, 85%) for ITO, TiO$_2$, NTO thin films respectively deposited on glass substrate and $T_{ave}$ is (85%, 81%, 82%) for ITO, TiO$_2$, NTO respectively for thin films deposited on PET substrate. The maximum value of transmittance at the optical wavelength of 550 nm was (86%, 91%, 85%) for thin films deposited on glass substrate and (85%, 81%, 81%) for ITO, TiO$_2$, NTO thin films deposited on PET substrate. In the same time the maximum value of the reflectance in the same wavelength range can be clearly observed with (9.6%, 8%, 11%), (12%, 14%, 14%) for ITO, TiO$_2$, NTO deposited on glass and PET substrates respectively (the inset of figure 2a and b) meaning that we note a high optical transmittance and a low reflectance in the visible region.

The transmittance and reflectance curves show no oscillation in the visible range. These thin films indicate low thicknesses, while we obtained a sharp edge between 350 nm and 400 nm because of the interband absorption edge.

Depending on the measured transmission and reflection spectra, the absorption coefficient $\alpha$ was obtained using the following equation:

$$\alpha(\lambda) = \frac{10^4}{d} \log_{10}\left(\frac{(1-R(\lambda))^2}{T(\lambda)}\right),$$

where $R$ is the reflectance, $T$ is the transmittance, and $d$ is the thickness of the sample.

The values of optical the energy gap $E_g [eV]$ are calculated by the Equation (2) of Tauc relation [69]

$$\alpha = \frac{A(E-E_g)^{\frac{1}{m}}}{E} \qquad (2)$$

where $\alpha$ is the absorption coefficient, $A$ an independent constant, $E_g$ the optical energy gap, $E$ the photon energy and $m = 0.5, 1.5, 2, 3$. These numbers refer to allowed direct transition (DT), forbidden direct transition (FDT) or allowed indirect transition (DIT), forbidden indirect transition (FIDT)) respectively. The figure of $(\alpha E)^{\frac{1}{m}}$ versus $E\ [eV]$ allows calculating the optical band gap by extrapolating the linear portion of the curve plotted to the energy axis $E$ where $(\alpha E = 0)$ in this graph. The values of $E_g$ due to these electronic transitions of ITO thin films deposited on glass and PET substrates are shown in the figure 3.

The same procedure is applied to obtain the values of energy gap for TiO$_2$, NTO thin films deposited on glass and PET substrates as illustrated in the table 1.

The extinction coefficient $k$ of the ultrathin films can be calculate by $\frac{\alpha\lambda}{4\pi}$, where $\alpha$ is the absorption coefficient, $\lambda$ is the wavelength of the light.

The figure 4 shows the extinction coefficient $k$ of these thin films. $k$ is very small in the visible range; indicating that ITO, TiO$_2$, NTO thin films are highly transparent. At $\lambda_{550}$ nm, the values of $k$ of the thin films are $(3.3 \times 10^{-2}, 2.1 \times 10^{-2}, 2.9 \times 10^{-2})$ and $(3.5 \times 10^{-2}, 4.6 \times 10^{-2}, 4.5 \times 10^{-2})$ for these thin films deposited on glass and PET substrates respectively.

It can be seen that the values of $k$ for films deposited on glass are smaller than that for PET substrates due to the absorption occurs on this substrate. $k$ value of TiO$_2$ is smaller than the others thin films in 550 nm while the $k$ values of ITO is higher than that for NTO for the same wavelength.

The refractive index can be determined using the following equation:

$$n = \left(\frac{1+R}{1-R}\right) - \sqrt{\frac{4R}{(1-R)^2} - k^2} \tag{3}$$

where $R$ is the reflectance and $k$ the extinction coefficient.

The figure 5 illustrates graphically the variations of the refractive index $n$ as a function of wavelength $\lambda$ for ITO, TiO$_2$, NTO thin films both deposited on glass and PET substrates respectively.

As it can be seen, the refractive index values $n$ decrease with the increasing of the wavelength $\lambda$ for the different films, the value of $n$ increases initially and then decreases. At $\lambda = 550$ nm the values of the refractive index of these thin films are 1.89, 2.06, 1.79 and 2.18, 1.97, 2.18 for these thin films deposited on glass and PET substrates respectively. The dispersion curve rises sharply towards the shorter wavelength, displaying the typical shape of a dispersion curve near an electronic interband transition.

The complex dielectric constants are calculated using equations 4 and 5 [70]

$$\varepsilon_r = n^2 - k^2 \quad (4)$$

$$\varepsilon_i = 2nk \quad (5)$$

The real and imaginary parts of the dielectric constants of ITO, TiO$_2$ and TiO$_2$:Nb deposited on glass and PET substrates are shown in the figure 6. This figure showed that $\varepsilon_r$ and $\varepsilon_i$ of the films decrease with the increasing of wavelengths. These curves allow extracting the dielectric constant of the thin films corresponding to the refractive $n$ and extinction $k$ coefficients.

### 3.2.2 (SE) Spectrophotometric Ellipsometry analyses

The ellipsometry measurements $\psi$, $\Delta$, $I_s$ and $I_c$ as well as their data fitting for ITO, TiO$_2$, NTO thin films deposited on glass and PET substrates using new amorphous experimental and theoretical (calculated) model are observed in the figure 7. They shows a good agreement between the experimentally and theoretically values of ($\psi, \Delta$) and ($I_s, I_c$) observed for films deposited on glass and PET substrates respectively. The physical parameters of the model are shown in the table 2.

The values of the thickness in (nm) of ITO, TiO$_2$, NTO was measured using a Dektak 6M profilometer as shown in the table 2. The thickness values of the thin films give a very good agreement compared to the values of the thickness of thin films measured by SE. One can observed that the difference of the values of thickness of the thin film thickness measured by two techniques (Dektak and SE) is less than one. This means that SE measurements have a good accuracy to measure the thickness of the thin films deposited on glass and PET substrates. Additionally the values of the optical constants of the thin films deposited on glass and PET substrates were measured and compared with the theoretical results obtained from the model.

The fit curves are determined by minimizing $\chi^2$(MSE) as closely as possible. To confirm the optical constants of ITO, TiO$_2$, NTO ultrathin films based on new amorphous model, the transmission and reflection spectrum are calculated between 200 nm and 2200 nm as shown in figure 8 considering the optical transmission spectra of ITO, TiO$_2$ and NTO thin films deposited on glass (figure 8 a) and PET (figure 8 b) substrates. The maximum value of transmittance in the optical wavelength $\lambda = 550$ nm was about [(85%, 83%), (88%, 88%), (88%, 87%)] for the thin films deposited on glass substrate and [(81%, 85%), (81%, 76%), (79%, 77%)] for the ITO, TiO$_2$, NTO thin films deposited on PET substrate respectively. In the same time; the maximum value of the reflectance in the same wavelength value can be clearly observed as [(10%, 12%), (10%, 10%), (11%, 12%)] and [(16%, 14%), (17%, 18%), (16%, 18%)] for ITO, TiO$_2$, NTO deposited on glass and PET substrates respectively are shown in the inset of figure 8 a and b. As a result, we observe a low reflectance in the visible range due to high optical transmittance.

According to the transmission and reflection spectrum results that have been obtained from the new amorphous model, the optical constants such as $n, k, E_g, \varepsilon_r, \varepsilon_i$ of TiO$_2$ thin film deposited on glass and PET substrates were calculated experimentally and theoretically as shown in figure 9. The same procedure is used to obtain the results of the optical constants for ITO and NTO. For ITO and NTO thin films the maximum values of the optical constants at the 550 nm wavelength are illustrated in the table 3.

According to the figure 9 and the table 3, the dispersion curves for the optical constants such as refractive index $n(\lambda)$ and the extinction $k(\lambda)$ coefficients, corresponding to the experimentally and theoretically real and imaginary $\varepsilon_r, \varepsilon_i$ dielectric constants were measured by ellipsometry analyses in the range of (200-2200) nm and compared with thin films deposited on glass and PET substrates. One can notice that the refractive index is higher in the case of niobium-doped titanium dioxide and in general in the case of NTO deposited on glass and PET, while the extinction index is higher for ITO deposited on glass and PET. The dielectric real constant is higher for NTO thin film and the imaginary dielectric constant is higher for ITO thin film. This proves that the real dielectric constant is a function of refractive index $n$ and the imaginary dielectric constant is a function of extinction coefficient $k$. The optical band gap values for different optical transitions according to the equation 2 for the films experimentally and theoretically are obtained by ellipsometry.

One can see an agreement between the measured values (EXP) of optical band gap for four types of electronic transitions of the ITO, TiO$_2$, NTO thin films measured by ellipsometry (SE) using a new amorphous model and the computed results (FIT) in the table 4.

### 3.2.3 Comparison the optical constants of thin films by Spectrophotometry and SE measurements

The agreement between the experimental and theoretical values calculated using the optical parameters found using both spectrophotometer and ellipsometry is illustrated in the figures 10, 11, 12 and tables (1, 4). The figure 10 shows the transmission and reflections curves of thin films deposited on glass and PET substrates at room temperature. In the figure 10, the $T_{ave}$ and $R_{ave}$ values of ITO, TiO$_2$ and NTO films deposited on glass substrate measured with the spectrophotometer is about (85% - 91%) and (8% - 10%) respectively. $T_{ave}$ and $R_{ave}$ for the same films deposited on glass substrate and measured by SE iss about (83% - 89%) and (10% - 12%) respectively. In addition $T_{ave}$ and $R_{ave}$ values of the same films deposited on PET substrate and measured by spectrophotometer are (81% - 85%) and (12% - 14%) respectively and $T_{ave}$ and $R_{ave}$ values of the same films deposited on PET substrate and measured by SE iss about (76% - 85%) and (14% - 17%) respectively.

In the figure 11 the $n_{ave}$ and $k_{ave}$ values of ITO, TiO$_2$ and NTO films deposited on glass substrate by the spectrophotometer are (1.79 - 1.97) and (0.02 - 0.03) respectively and $n_{ave}$ and $k_{ave}$ values of the same films deposited on glass substrate and measured by SE are (1.93 - 2.06) and (0.03 - 0.04) respectively. In addition; $n_{ave}$ and $k_{ave}$ values of the same films deposited on PET substrate and measured by spectrophotometer are (2.06 - 2.16) and (12% -14%) respectively and $n_{ave}$ and $R_{ave}$ values of the same films deposited on PET substrate and measured by SE were about (0.04 - 0.05) and (0.004 - 0.006) respectively.

The figure 12 illustrates the $\varepsilon_{r_{ave}}$ and $\varepsilon_{i_{ave}}$ values of ITO, TiO$_2$ and NTO films deposited on glass substrate. We found for $\varepsilon_{r_{ave}}$ and $\varepsilon_{i_{ave}}$ (3.21 - 3.92) and (0.02 - 0.04) respectively measured with the spectrophotometer and $\varepsilon_{r_{ave}}$ and $\varepsilon_{i_{ave}}$ values of (3.74 - 4.25) and (0.003 - 0.04) respectively for the same films deposited on glass substrate and measured by SE. In addition, $n_{ave}$ and $k_{ave}$ values of the same films deposited on PET substrate and measured by spectrophotometer are (4.26-4.71) and (0.16-0.20) respectively and $\varepsilon_{r_{ave}}$ and $\varepsilon_{i_{ave}}$ values of the same films deposited on PET substrate and measured by SE are (4.84-6.04) and (0.17-0.30) respectively.

Finally the agreement between the experimental optical band gap values for films deposited on both glass and PET substrates measured by spectrophotometer and the optical band gap values for films deposited on both glass and PET substrates by SE for the four types of electronic transitions is shown in table 1 and 4 is good.

## 4. Conclusion

ITO, TiO$_2$ and NTO thin films with an average thickness of 20 nm were deposited on glass and PET substrates with a DC sputtering technique. The films present different particle sizes when they are deposited on glass substrate or when there are deposited on plastic substrates (PET). The effect of the substrate nature influences the TiO$_2$ and NTO films deposition. For ITO films deposition, the roughness of the films is smaller than the other thin films deposited on the same substrates. The optical constants of the films deposited onto glass and PET substrates gives an average transmission values in the visible region of about 85 - 91% for the films deposited on glass substrates and 81 - 85% for the PET substrates. These values of transmission make these films very suitable for optoelectronic applications. Determinations of the optical band gap values for four modes of electronic transitions of the films deposited onto glass and PET substrates give an excellent agreement experimentally and theoretically using different methods as UV and SE (with a new amorphous model). Moreover, the refractive index values for all the samples decrease rapidly and reach a constant at long wavelength range. A DC sputtering technique can be an excellent choice to obtain thin films with high optical transmission (wide band gap) and makes these films excellent candidates for optoelectronic devices, especially flexible solar cell applications.

**Acknowledgement**

Authors are grateful to Nicolas Mercier, Magali Allain for providing the necessary facilities for XRD studies, Also, to Jean-Paul Gaston and Celine Eypert from Jobin Yvon Horiba Company for the spectroscopic ellipsometry measurements and to Cecile Mézière, Valerie BONNIN for help with the chemicals and corresponding equipment's.

**Figures Captions:**

Figure 1: SEM images for deposited layers: ITO, $TiO_2$ and (NTO) on glass and PET substrates

Figure 2: Transmission and reflection spectra for the ITO, $TiO_2$, NTO thin films deposited on glass (a) and PET (b) substrates, respectively.

Figure 3: Variation of the optical band gap $E_g$ as a function of the incident photon Energy E for ITO thin film deposited on glass (a, b) and PET (c, d) substrates.

Figure 4: Variation of extinction coefficient k as a function of incident photon wavelength λ for ITO, $TiO_2$, NTO thin films deposited on glass and PET substrates.

Figure 5: Variations of the refractive index $n$ as a function of wavelength λ for ITO, $TiO_2$, $TiO_2$:Nb thin films each deposited on glass and PET substrates respectively.

Figure 6: Variation of the real part (a, b) and imaginary part (c, d) of dielectric constants of ITO, $TiO_2$, NTO thin films deposited on glass (a, c) and PET (b, d) substrates, respectively.

Figure 7: Ellipsometry spectra of experimental and theoretical (ψ, Δ) and ($I_s$, $I_c$) and best fit of ITO, thin film deposited on glass (a, b) and PET (c, d) substrates as a functions of λ with "New amorphous" dispersion formula.

Figure 8: Transmission and reflection spectra (EXP & FIT) for ITO, $TiO_2$, NTO ultrathin films deposited on glass (a) and PET (b) substrates, respectively, calculated by SE.

Figure 9: Dispersion curves of experimental (EXP) and theoretical (FIT) optical constants, (a, e) $(\alpha E)^2$ versus E and (b, f) refractive and extinction coefficients versus wavelength, (c, g) real and imaginary dielectric constants versus wavelength of $TiO_2$ ultrathin films with a new amorphous model deposited on glass and PET substrates respectively.

Figure 10: Transmission and reflection curves as a function of wavelength obtained by spectrophotometer and ellipsometry for ITO, $TiO_2$ and NTO thin films of about 20 nm thickness deposited on glass (a, b) and PET (c, d) substrates respectively.

Figure 11: Refractive and extinction coefficients curves as a function of wavelengths obtained by spectrophotometer and ellipsometry for ITO, TiO$_2$ and NTO thin deposited on glass (a, b) and PET (c, d) substrates respectively.

Figure 12: Dielectric constants real and imaginary curves as a function of wavelength obtained by spectrophotometer and ellipsometry for ITO, TiO$_2$ and NTO thin deposited on glass (a, b)and PET (c, d) substrates respectively.

# List of tables:

Table 1: shows the values of optical energy gap obtained using the values of $m$ for ITO, TiO$_2$, NTO thin films deposited on glass and PET substrates.

Table 2: New amorphous parameter values used for fitting the thickness and the optical constants (experiment and measured) of the ITO, TiO$_2$, NTO thin films deposited on glass and PET substrates.

Table 3: shows the maximum values of $n$, $k$, $\varepsilon_r$, $\varepsilon_i$ of thin films deposited on glass (a) and PET (b) substrates respectively.

Table 4: shows the energy gap values for four optical transitions of the films.

| Materials | | |
|---|---|---|
| Substrate | Glass | PET |
| ITO | 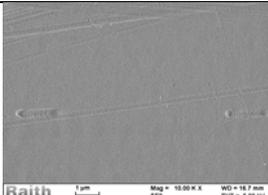 | 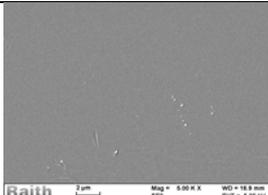 |
| TiO$_2$ | 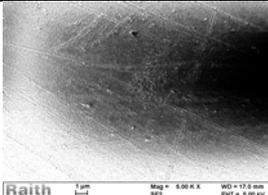 | 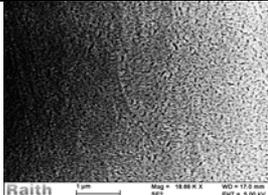 |
| TiO$_2$:Nb (NTO) | 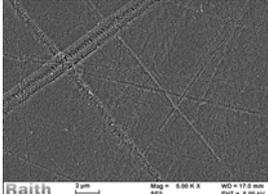 | 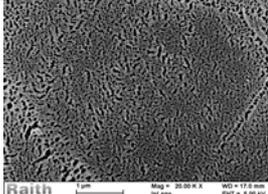 |

Figure 1

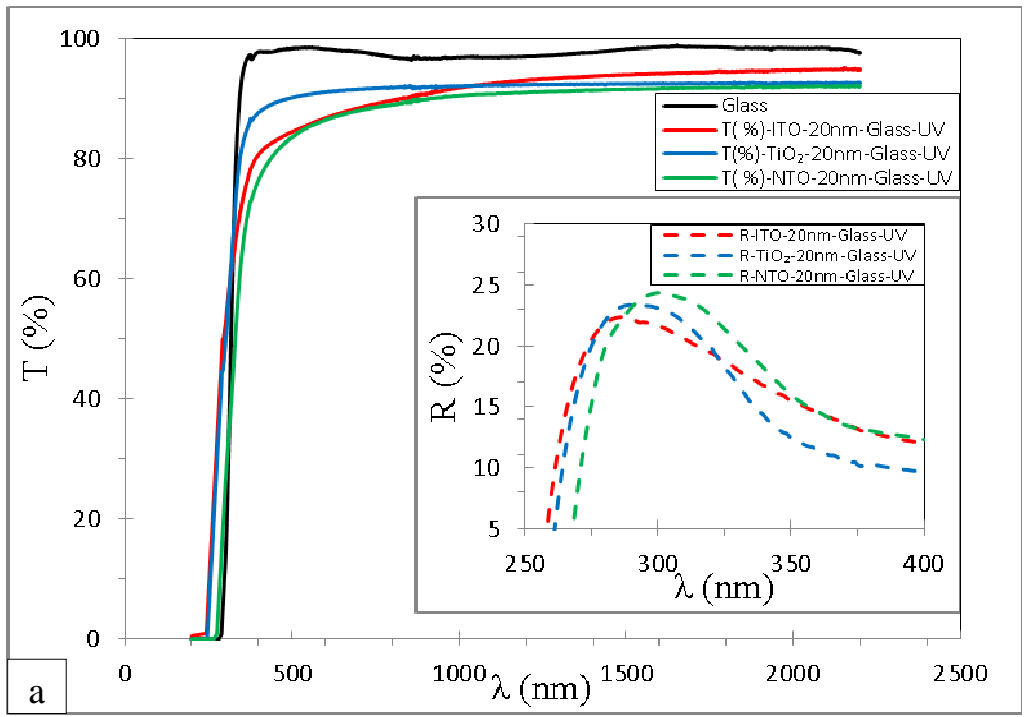

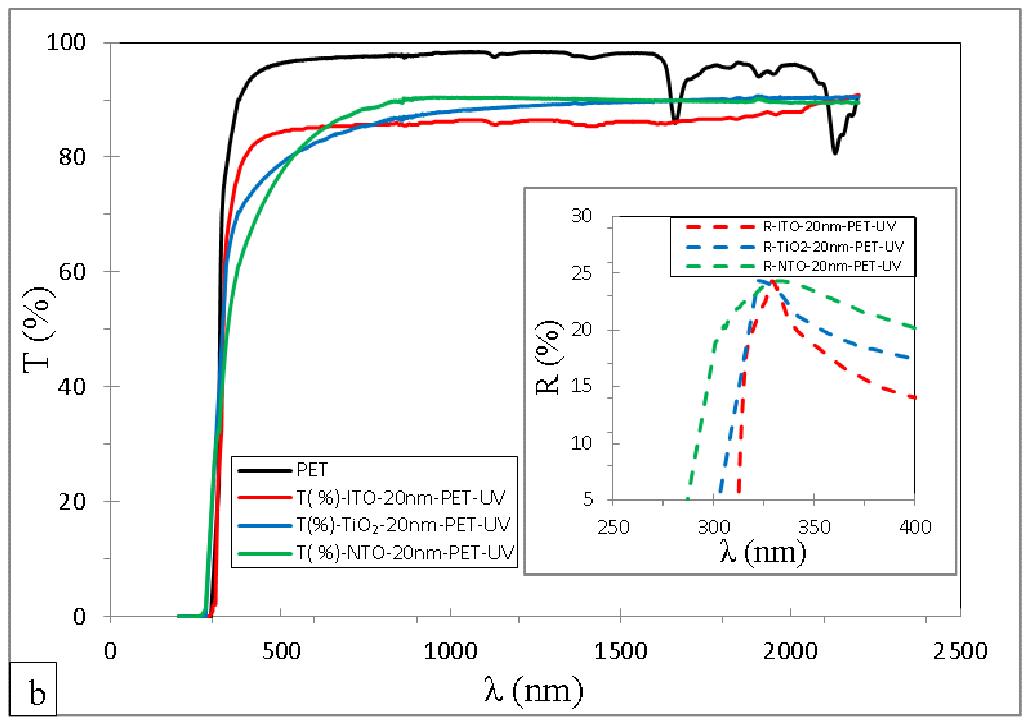

Figure 2

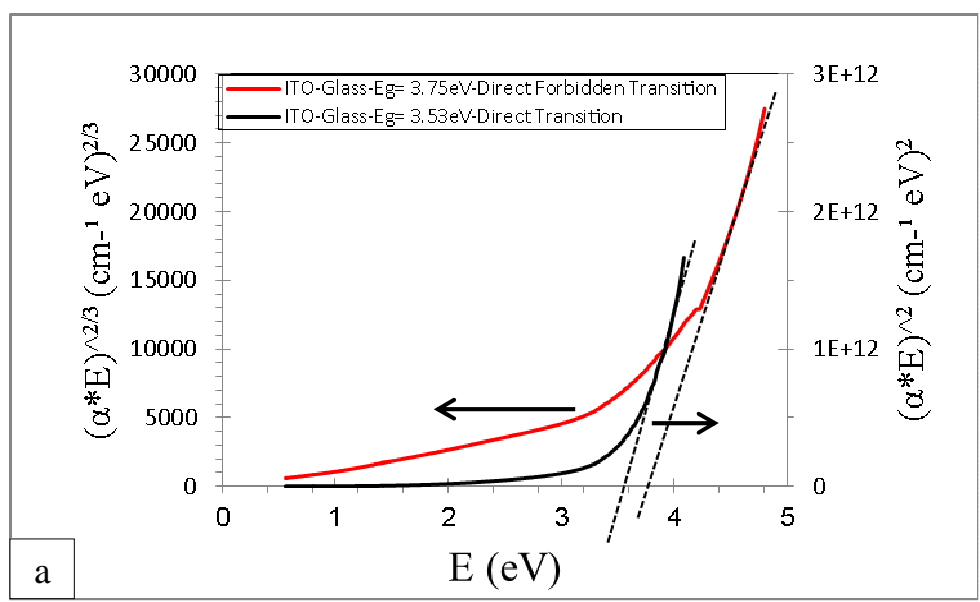

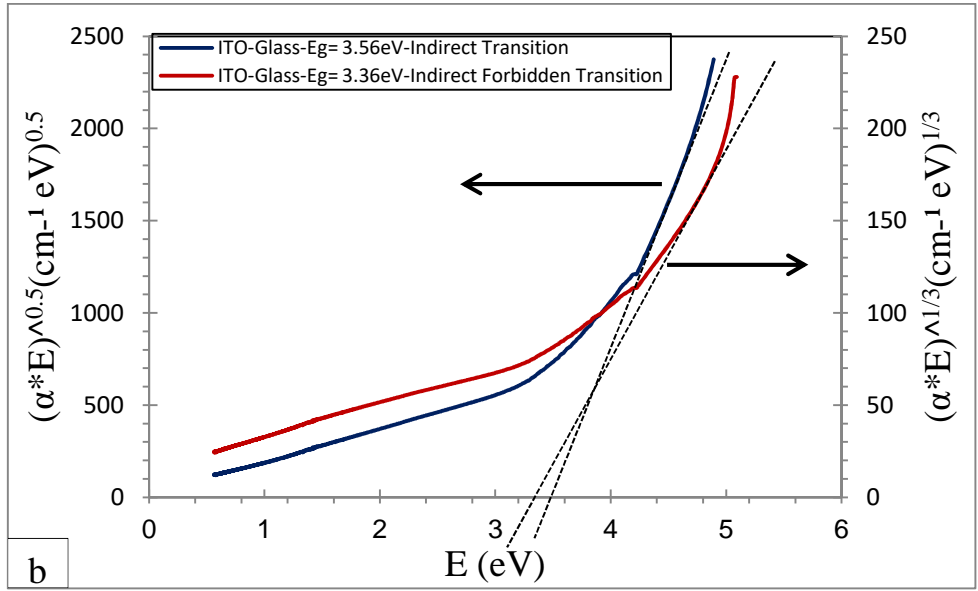

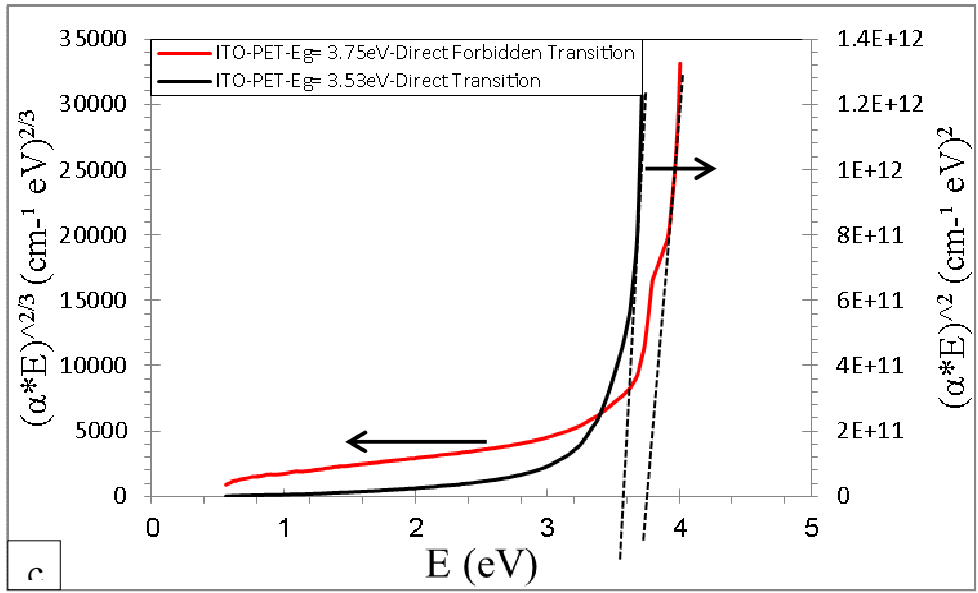

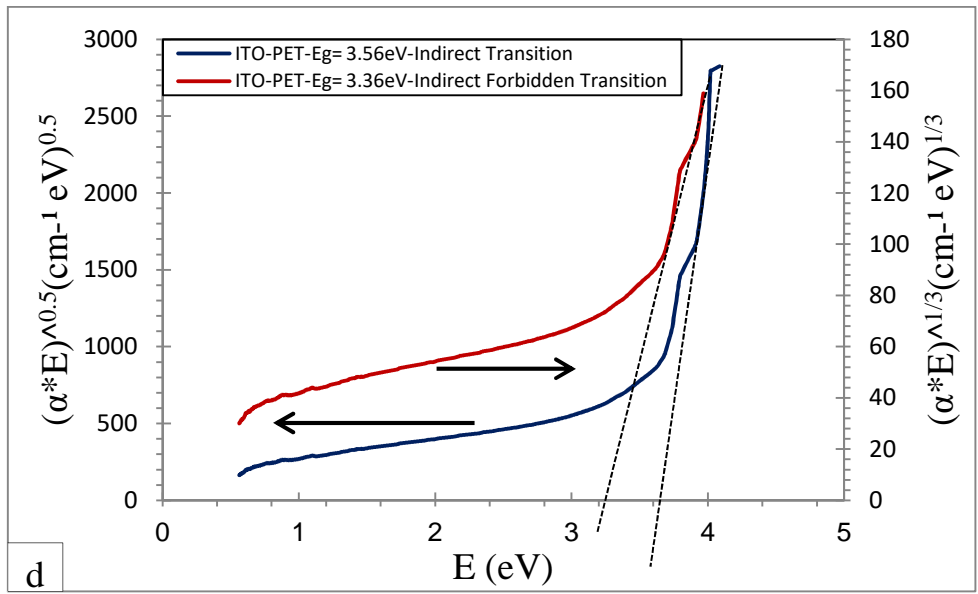

Figure 3

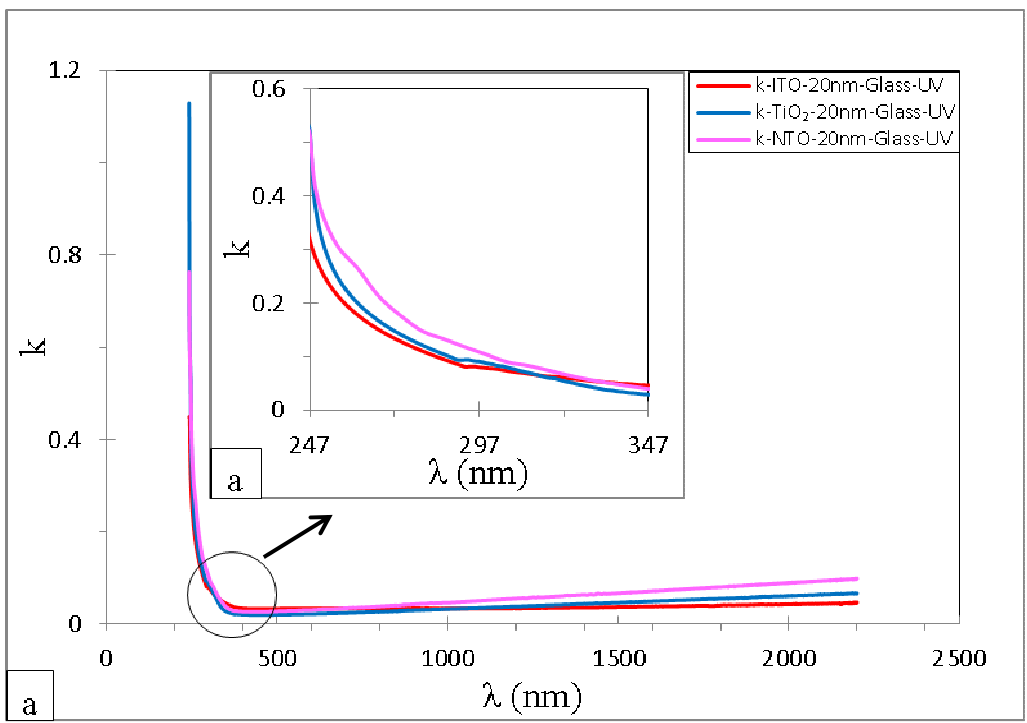

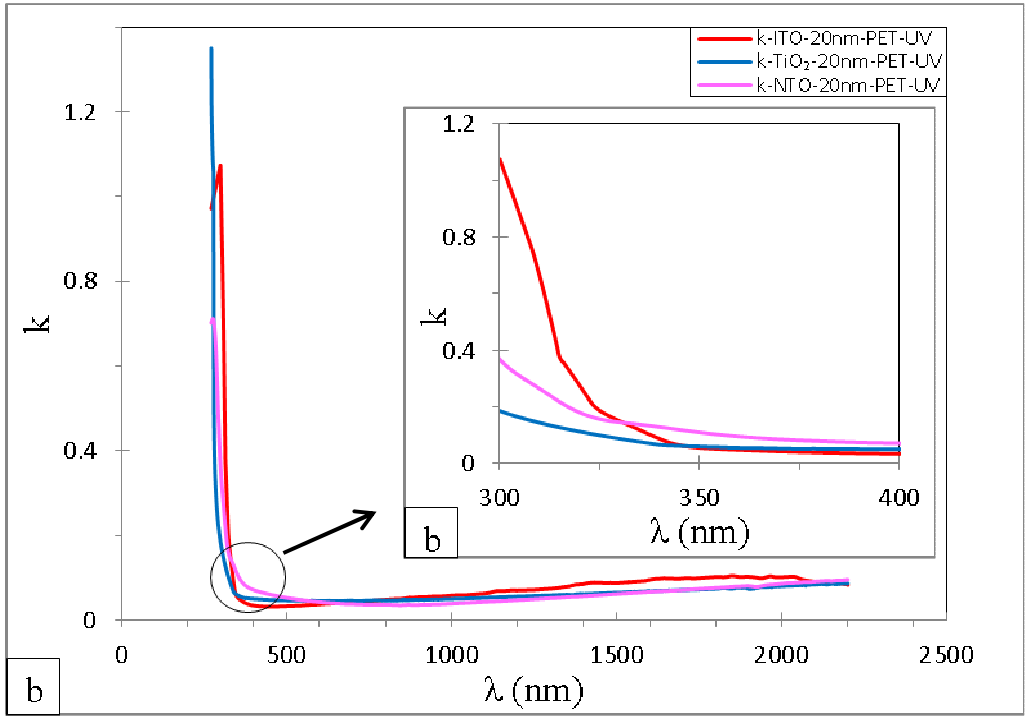

Figure 4

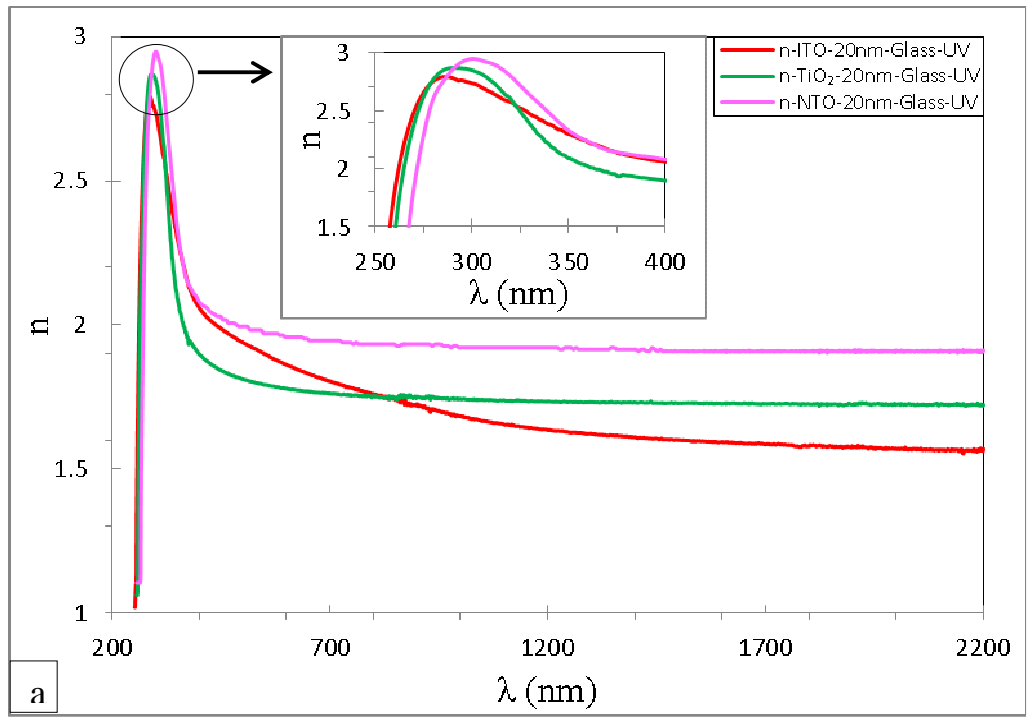

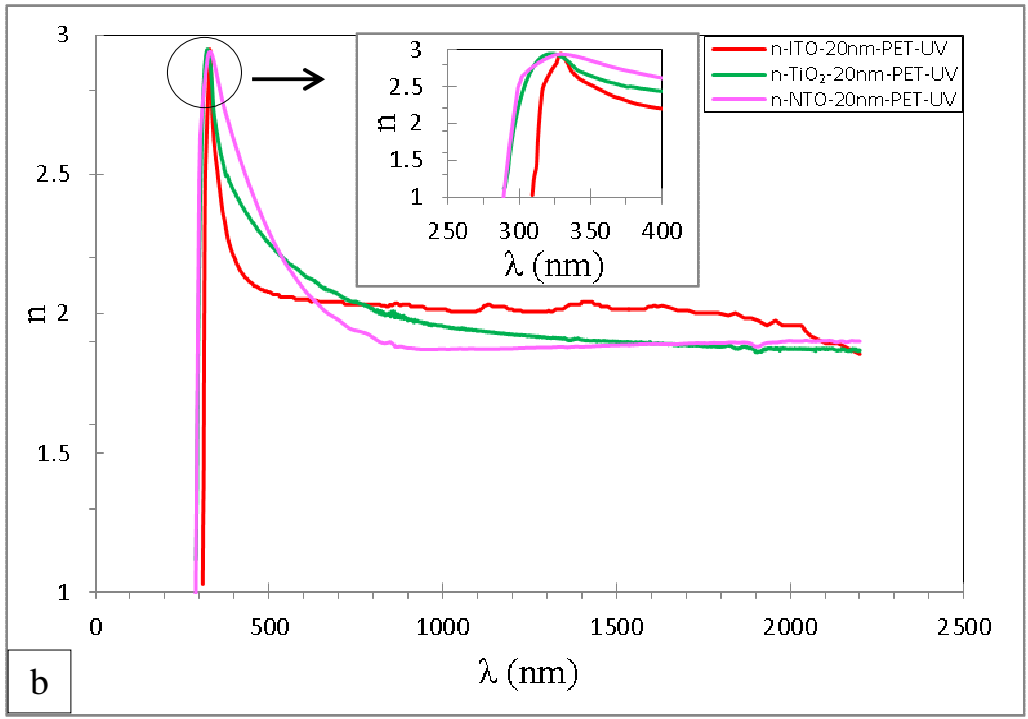

Figure 5

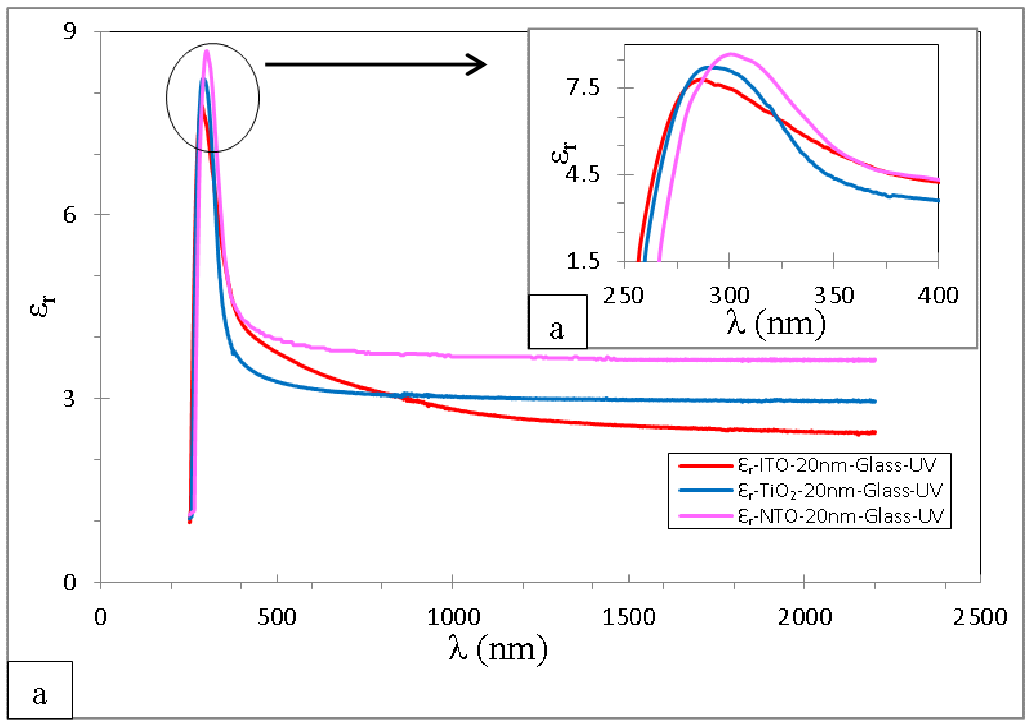

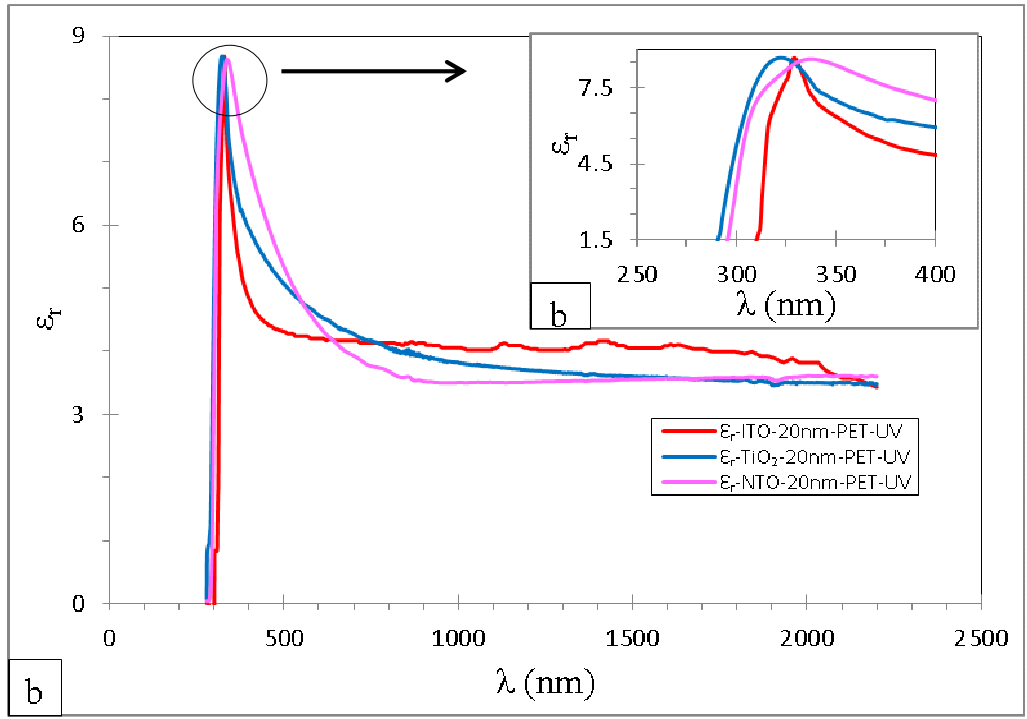

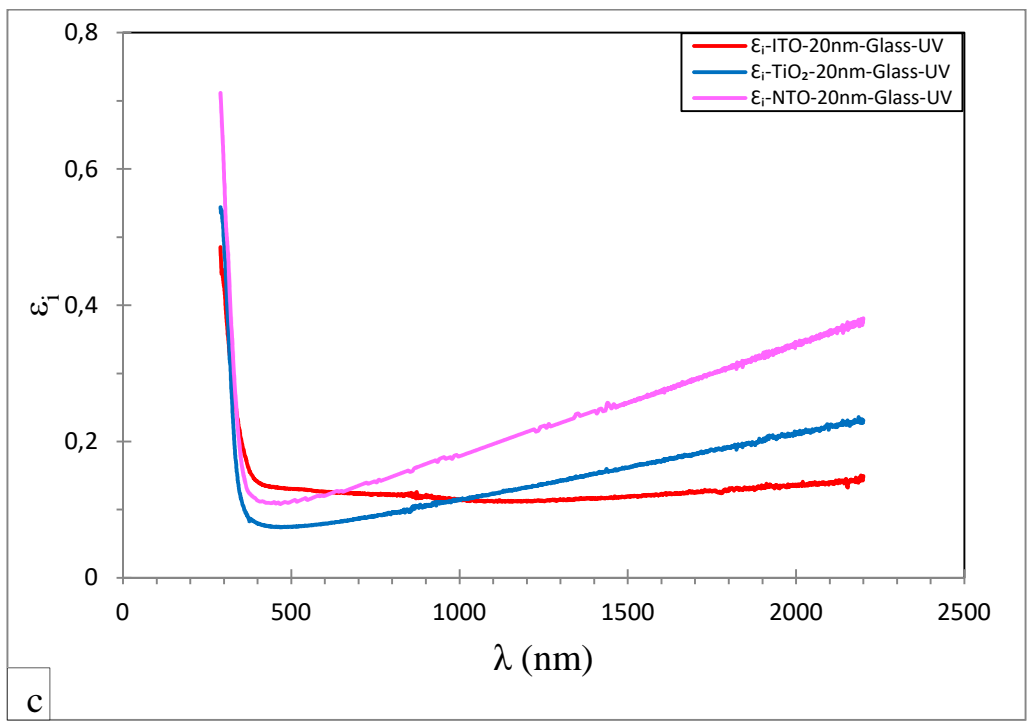

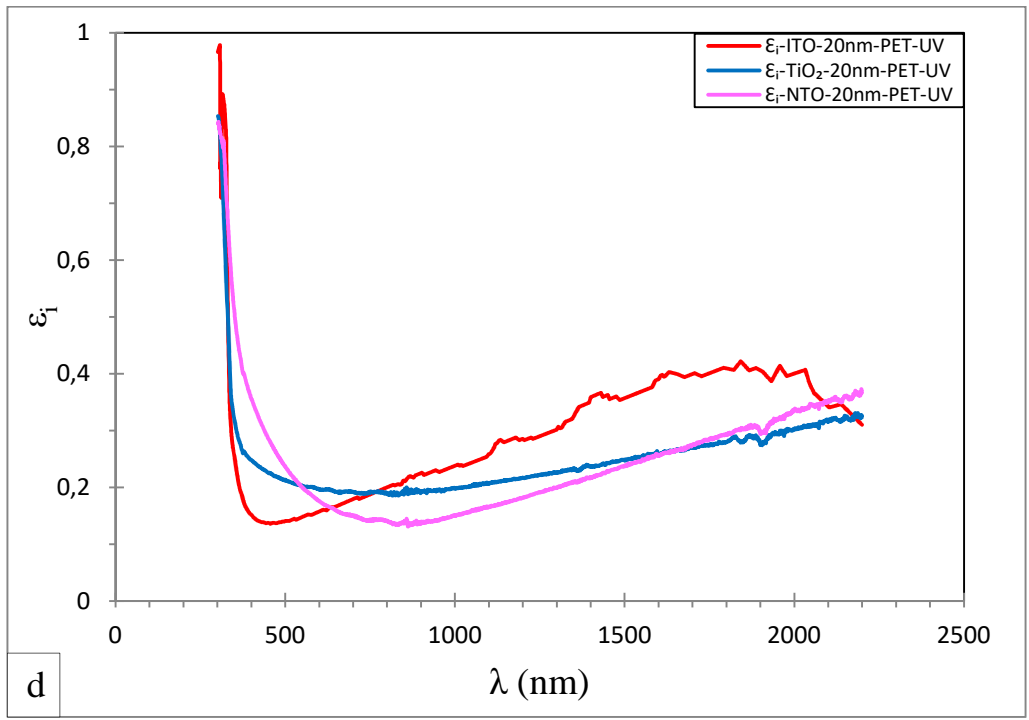

Figure 6

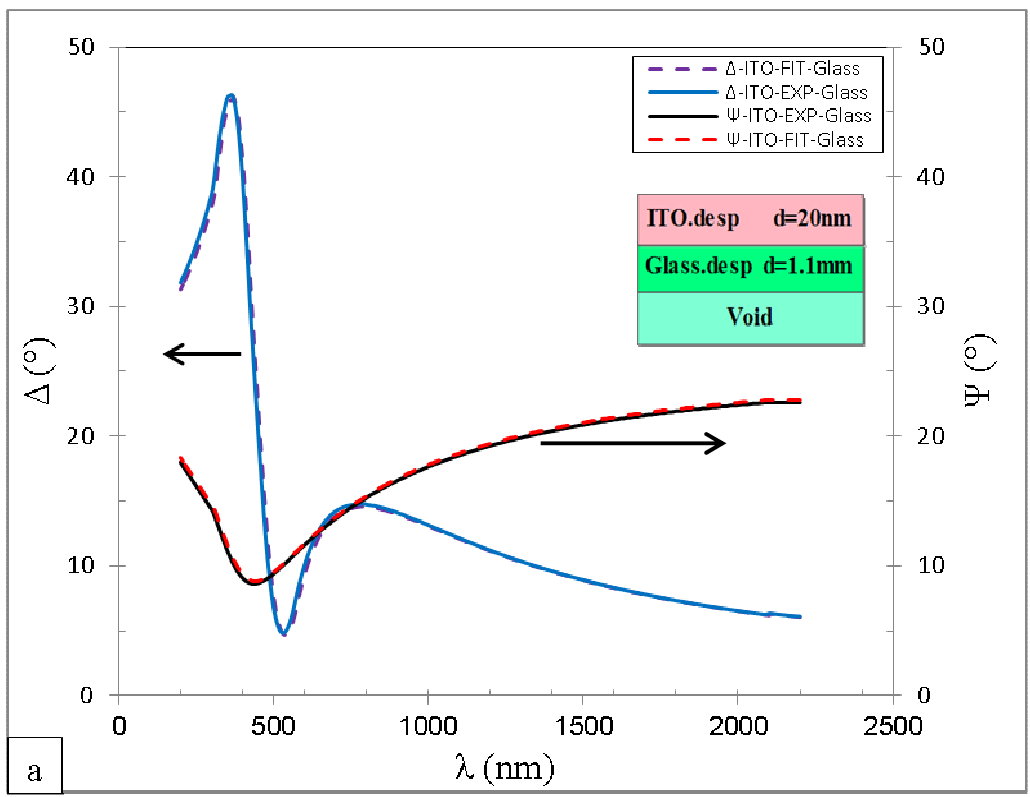

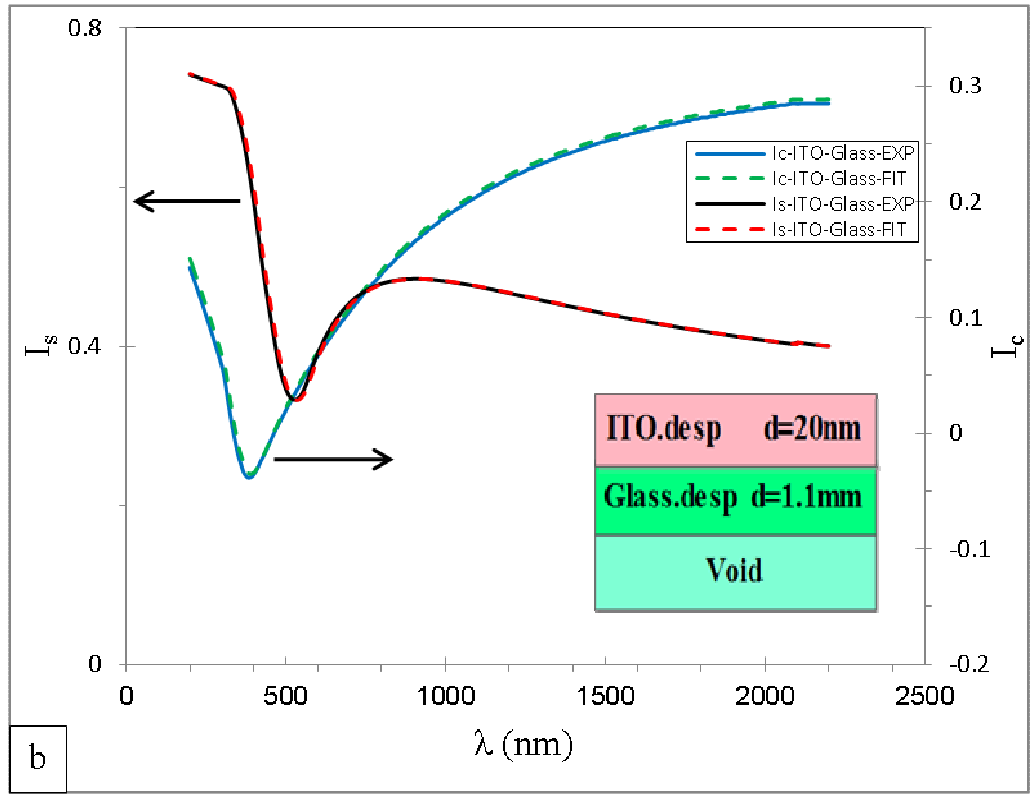

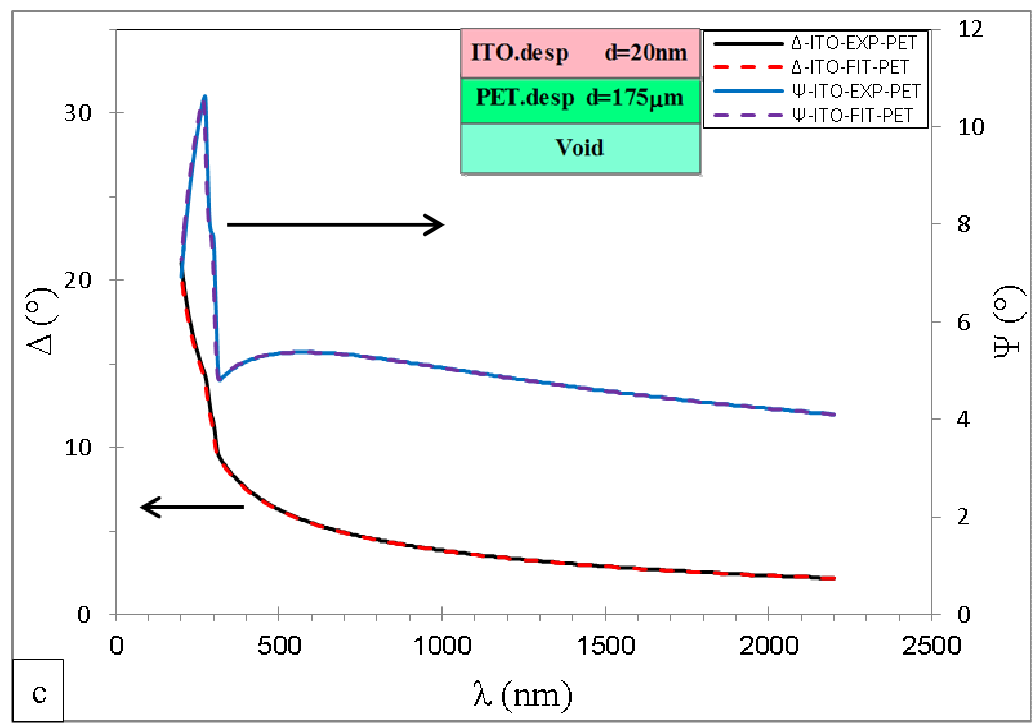

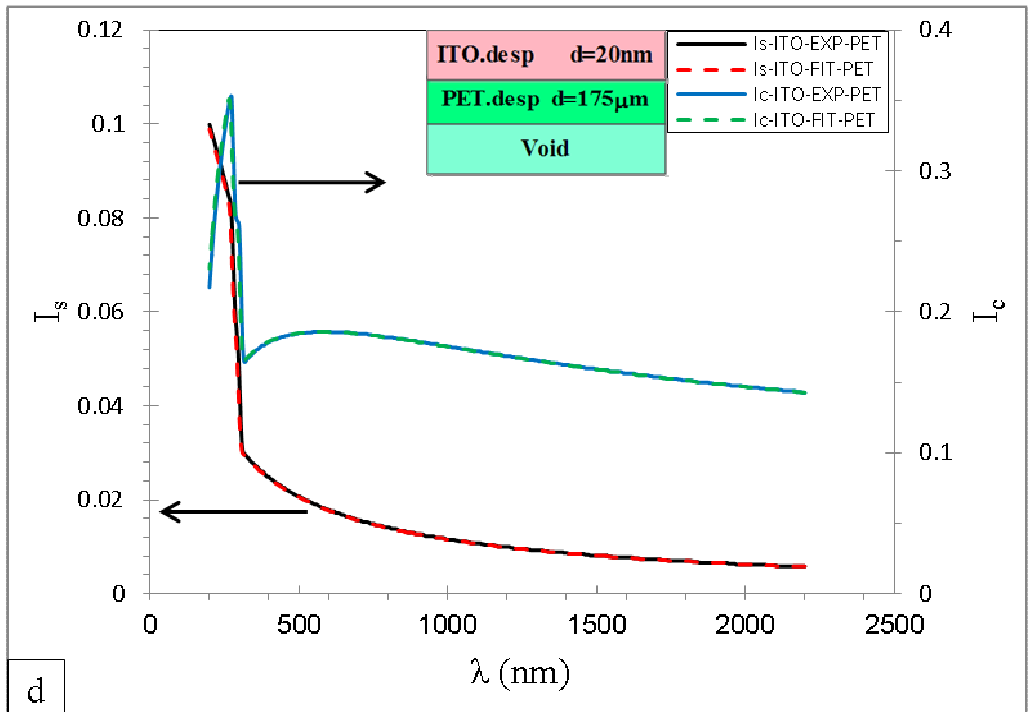

Figure 7

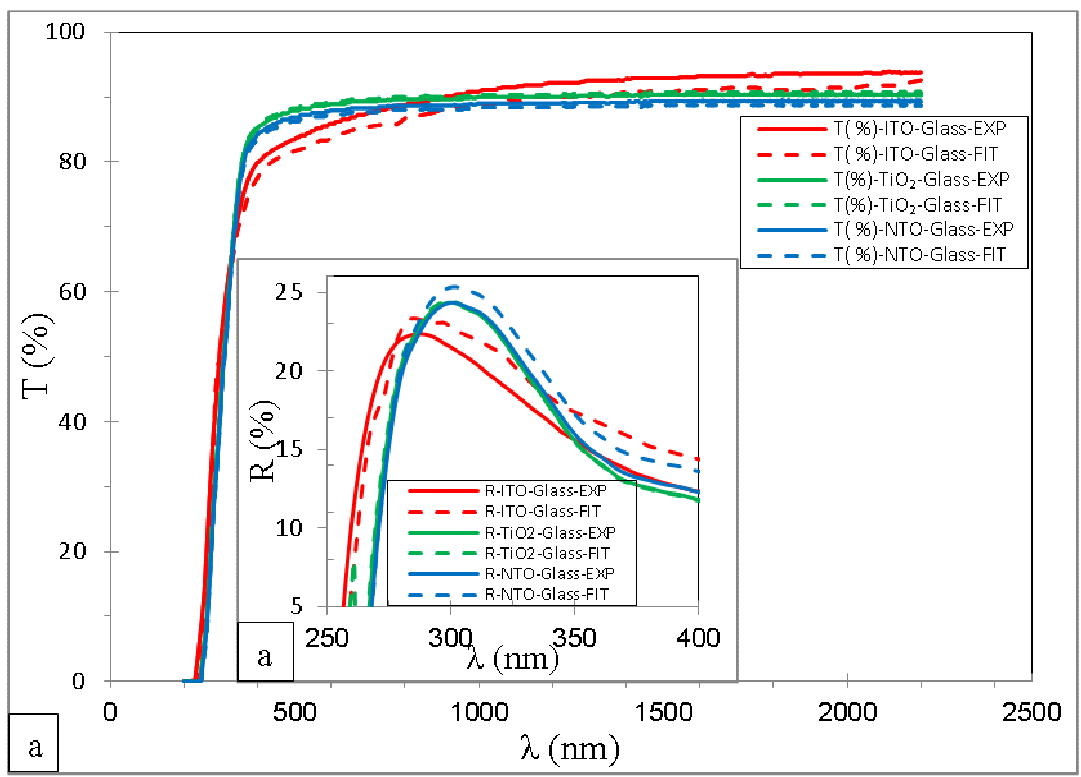

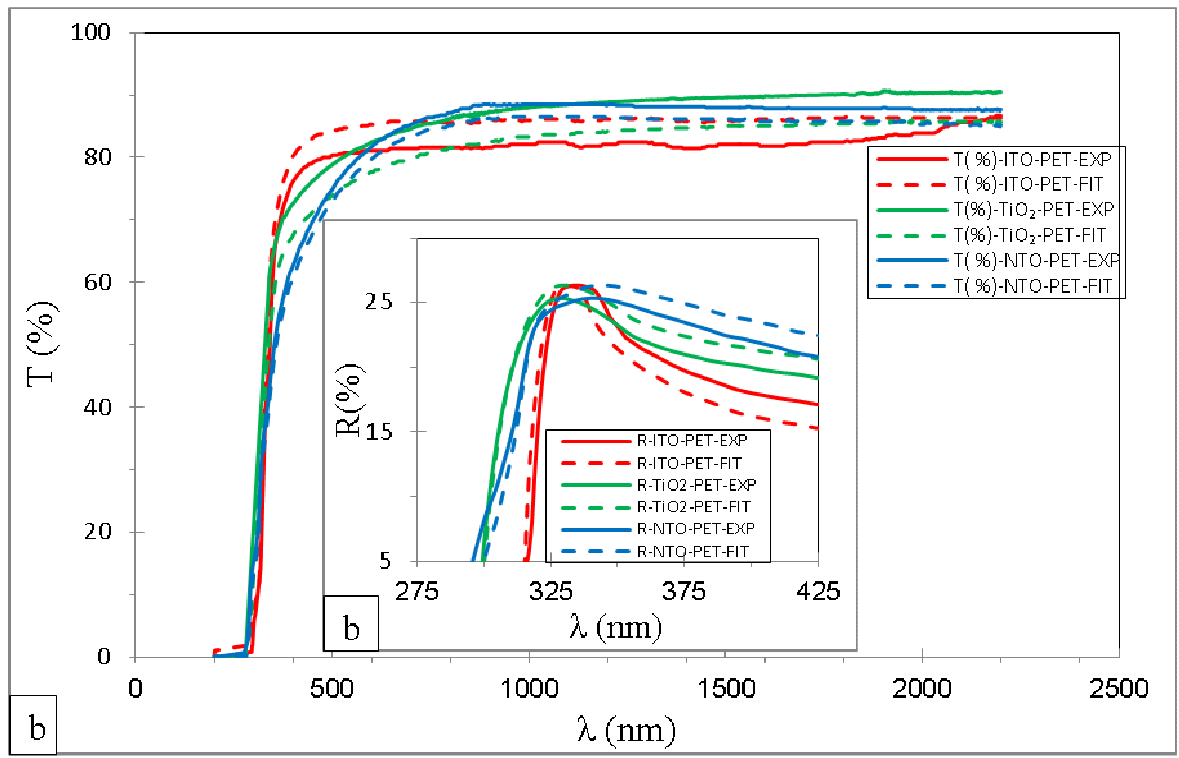

Figure 8

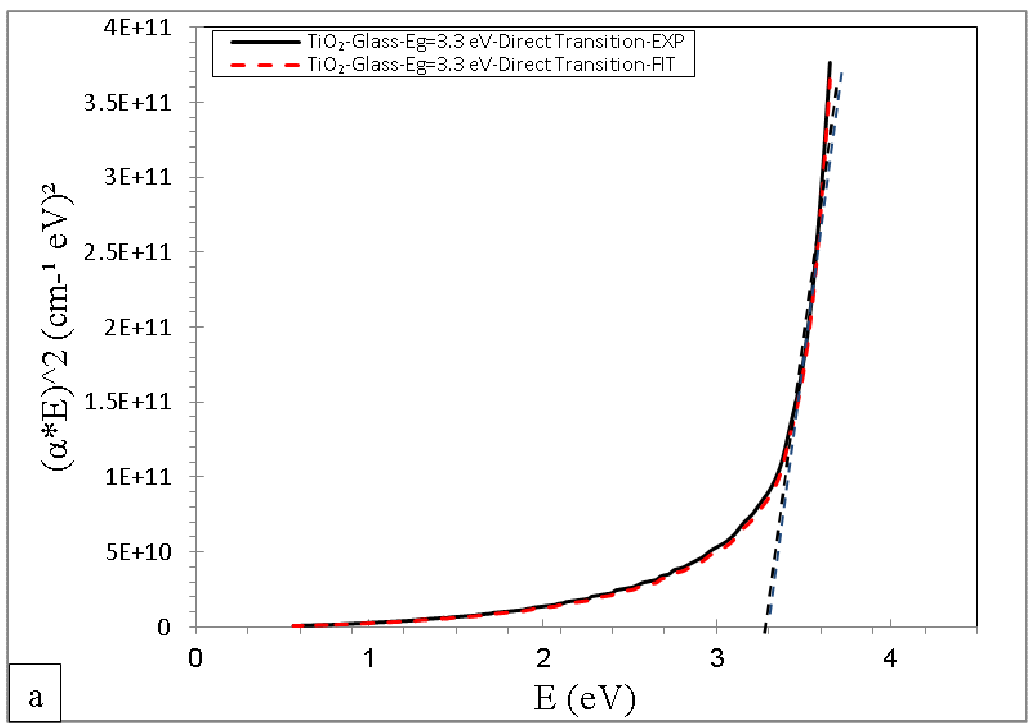

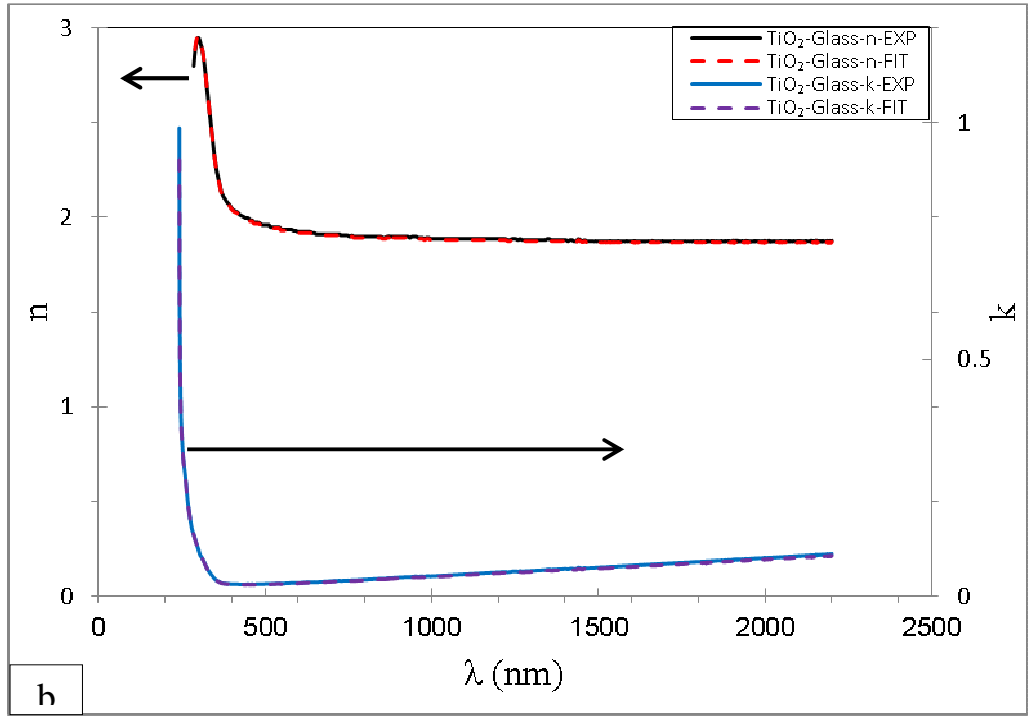

b

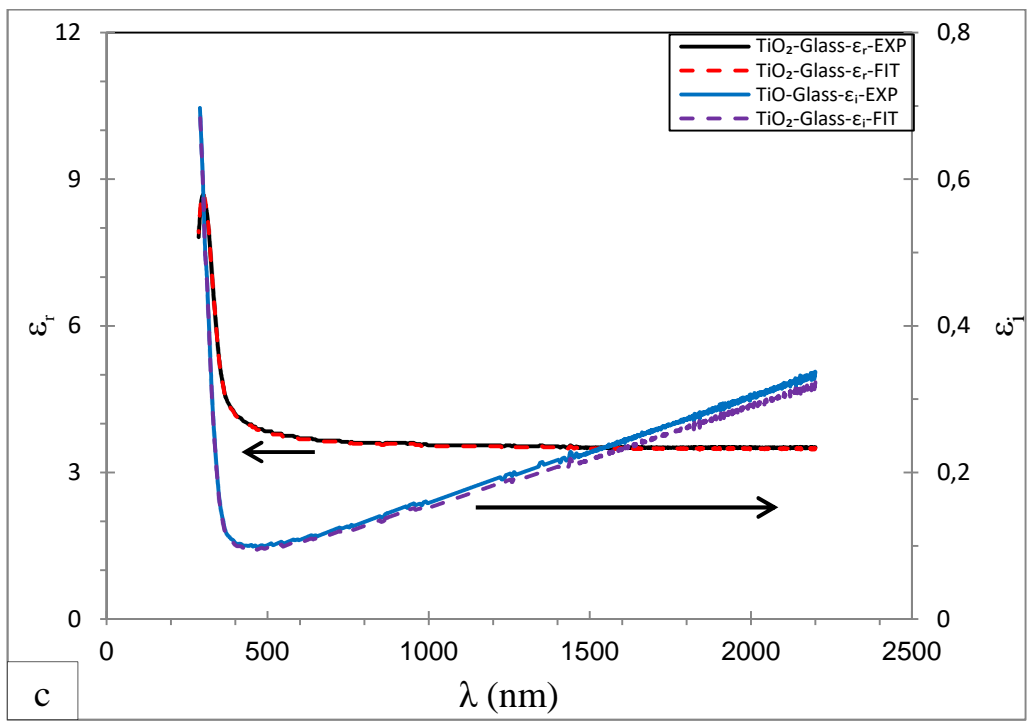

c

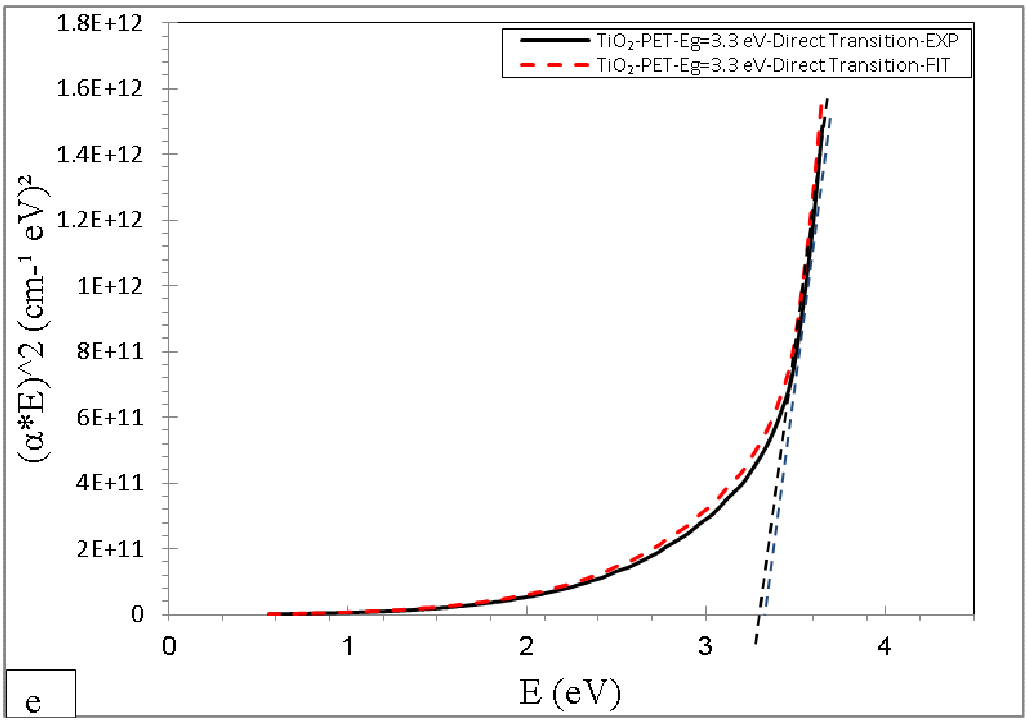

e

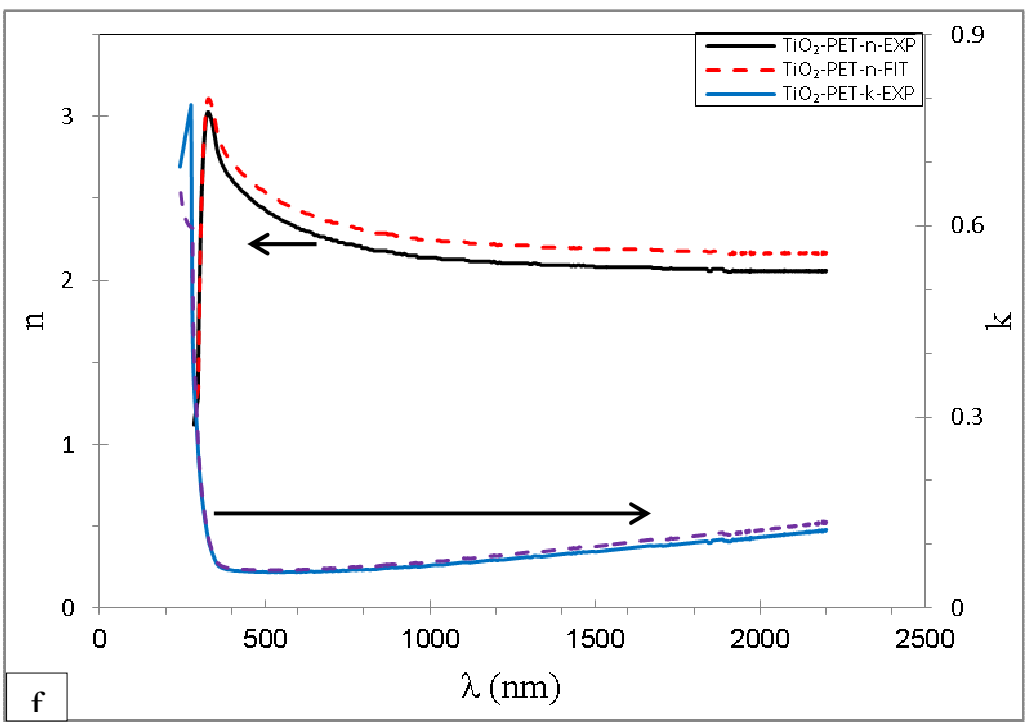

f

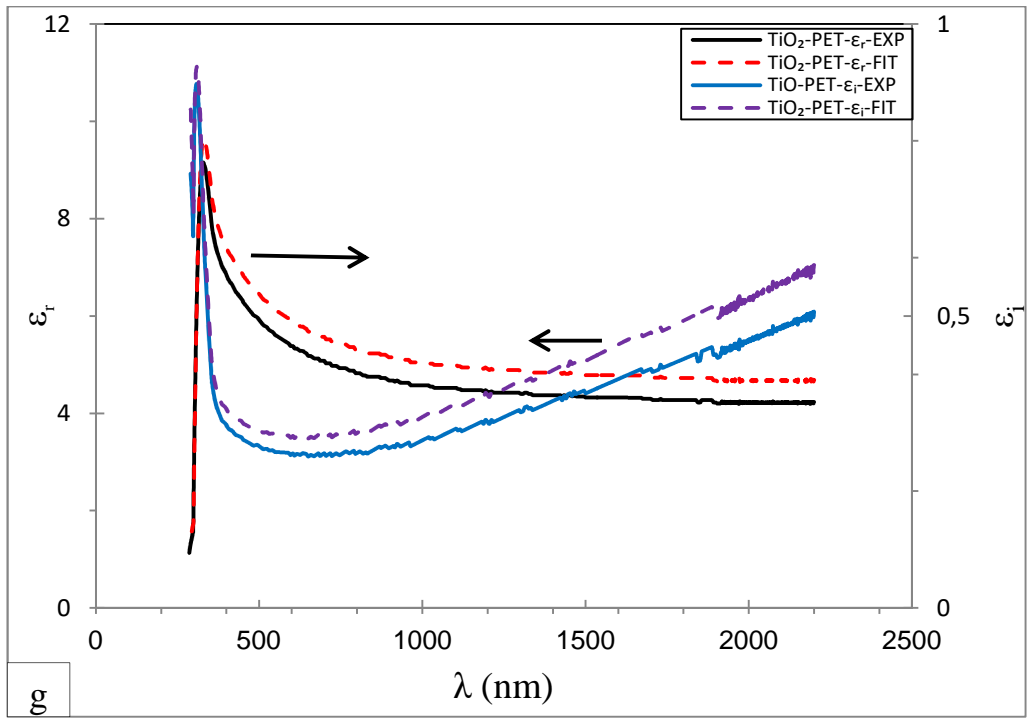

Figure 9

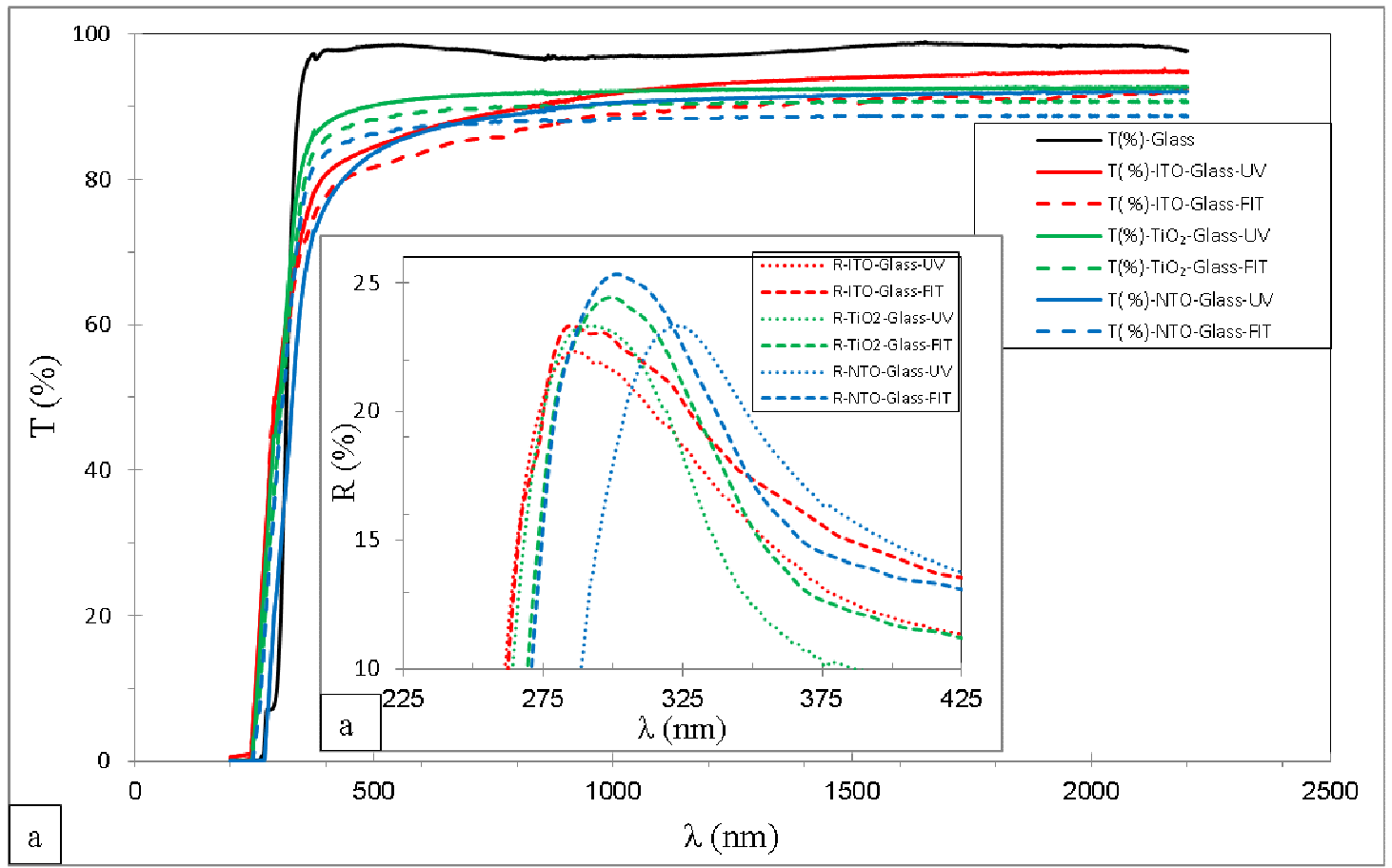

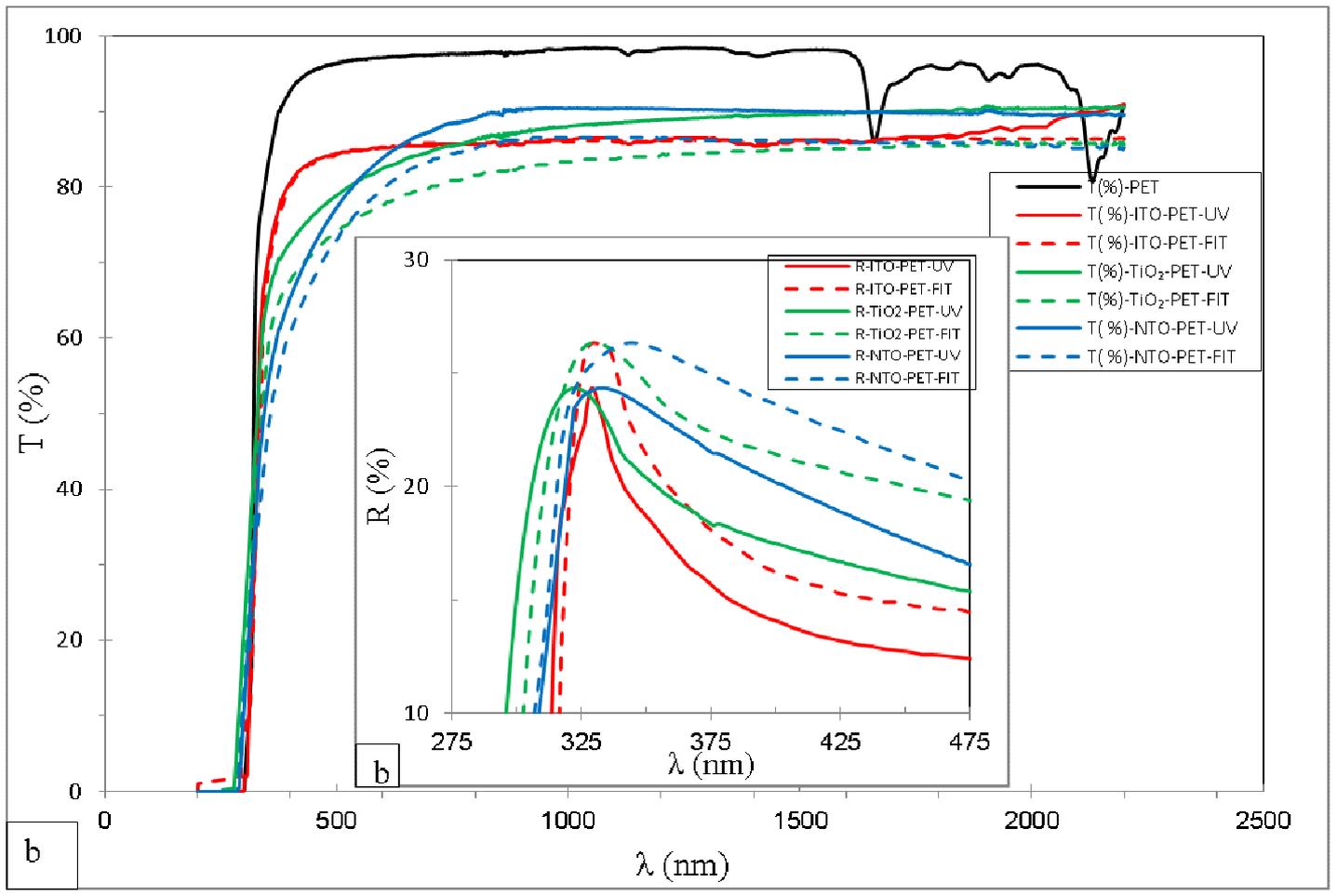

Figure 10

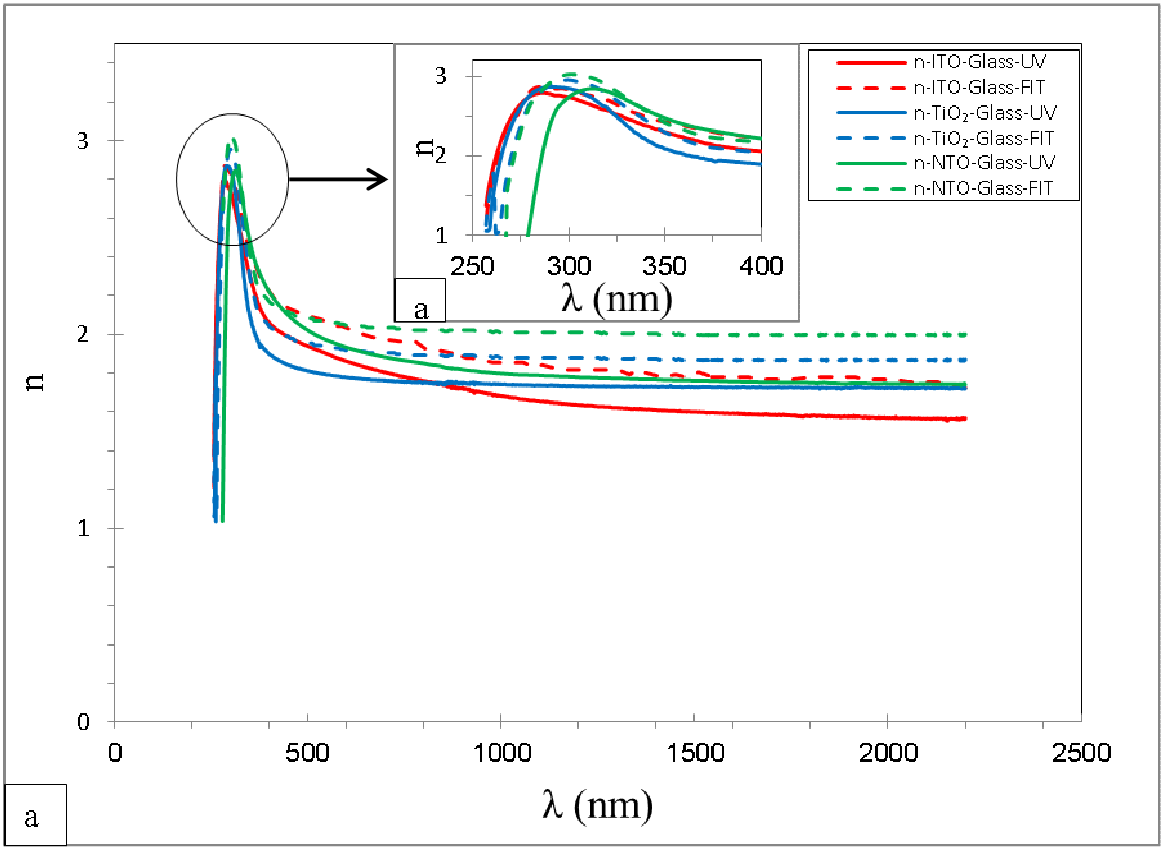

a

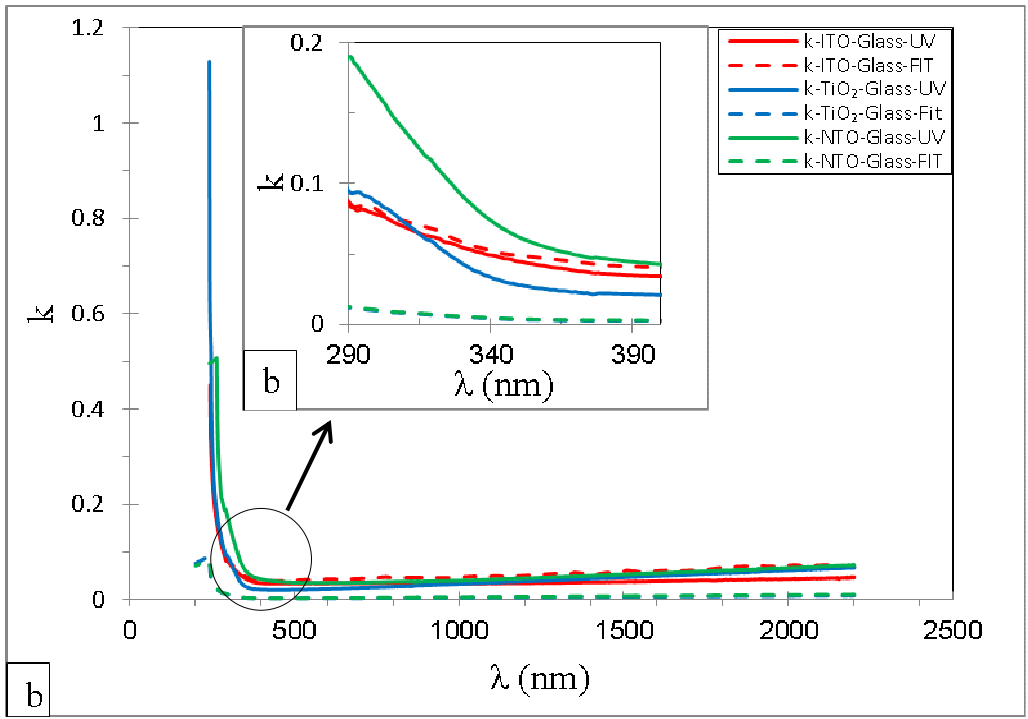

b

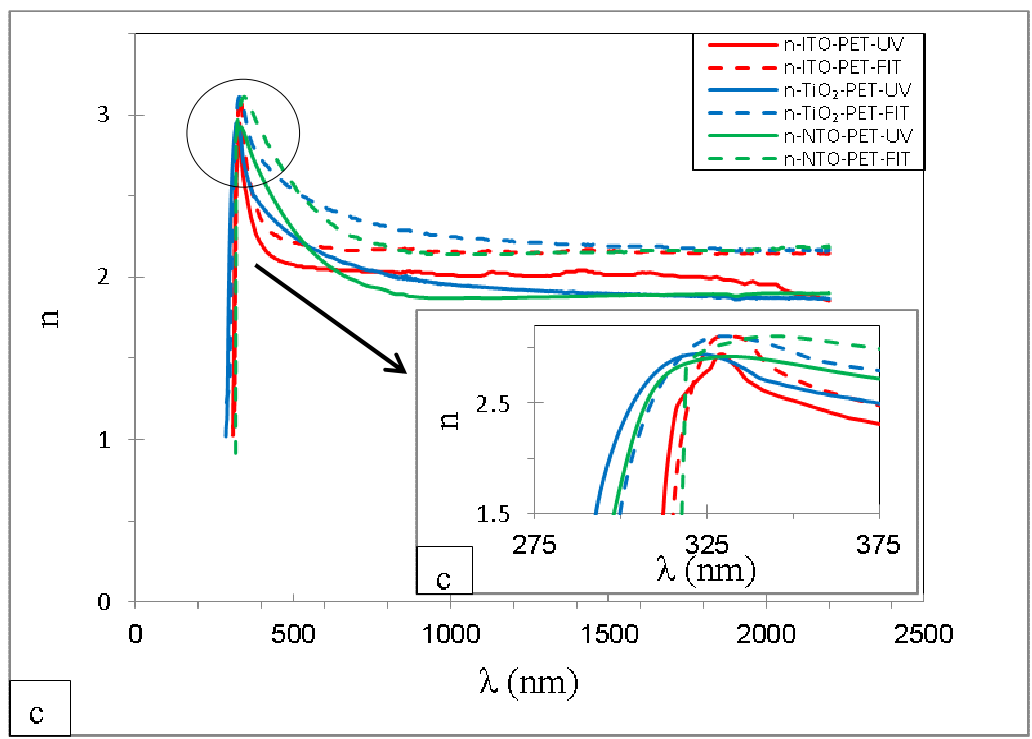

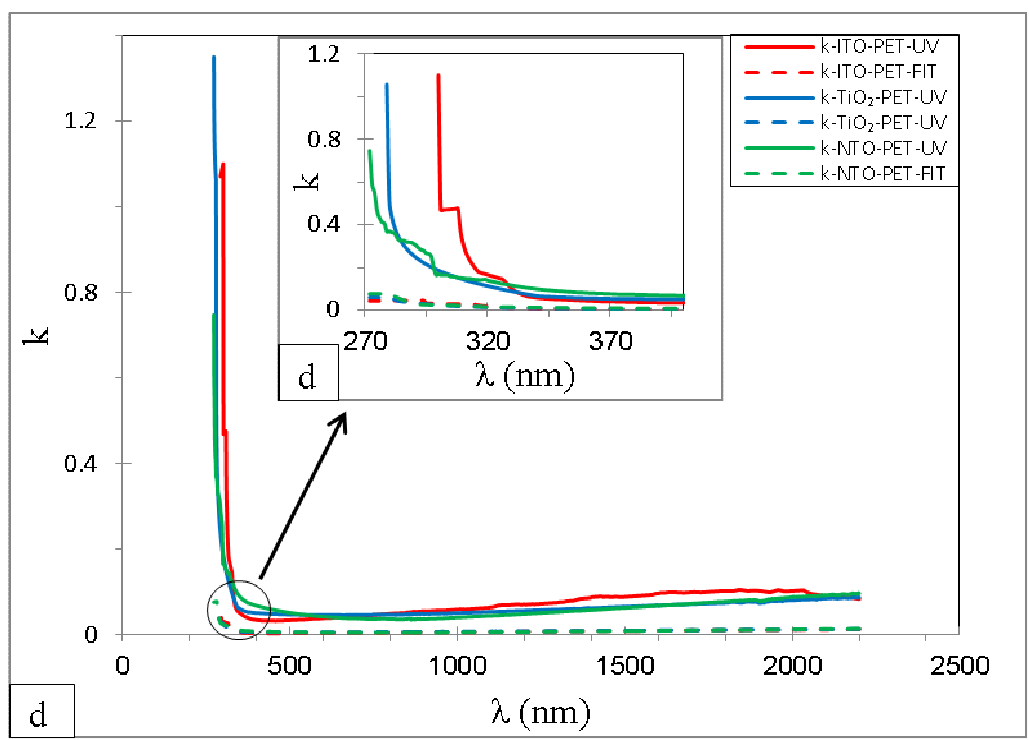

Figure 11

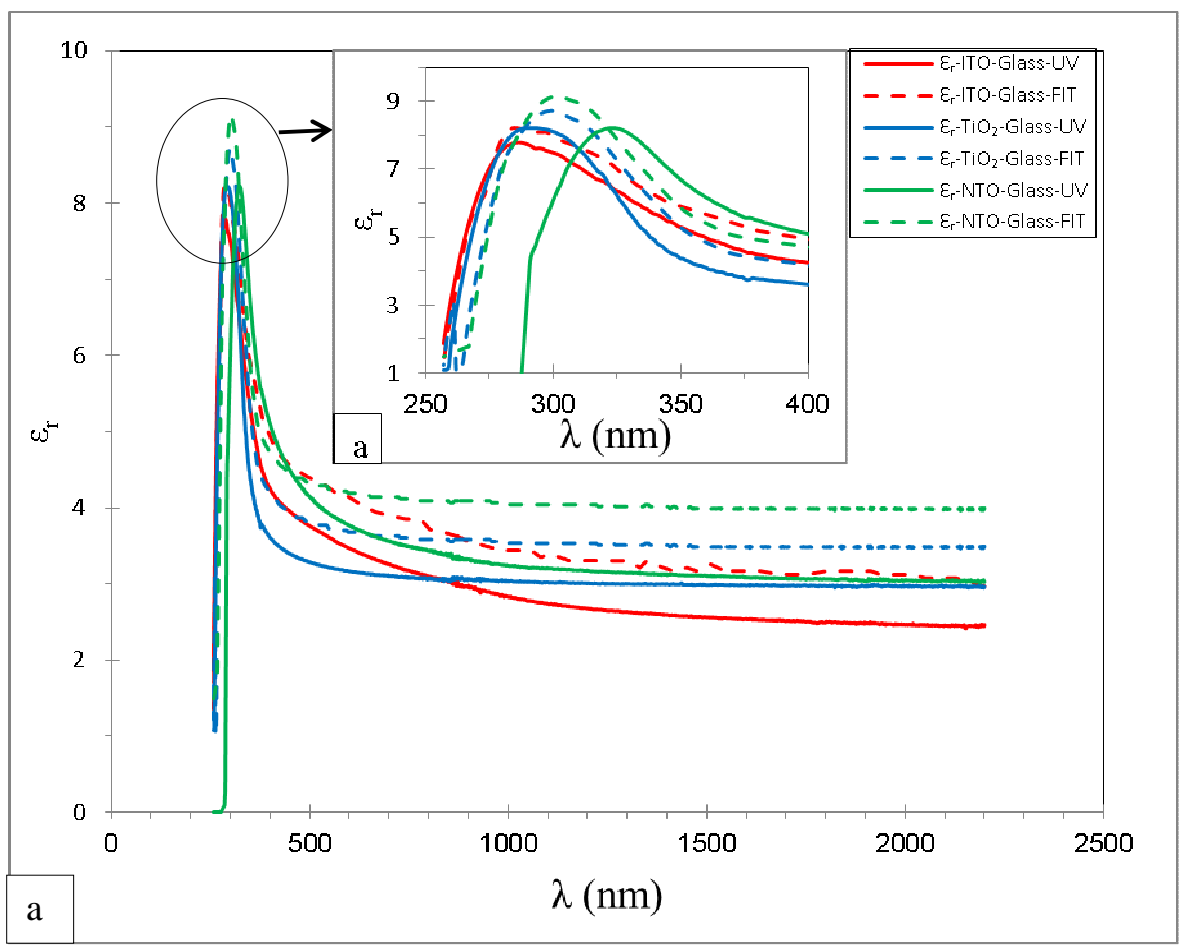

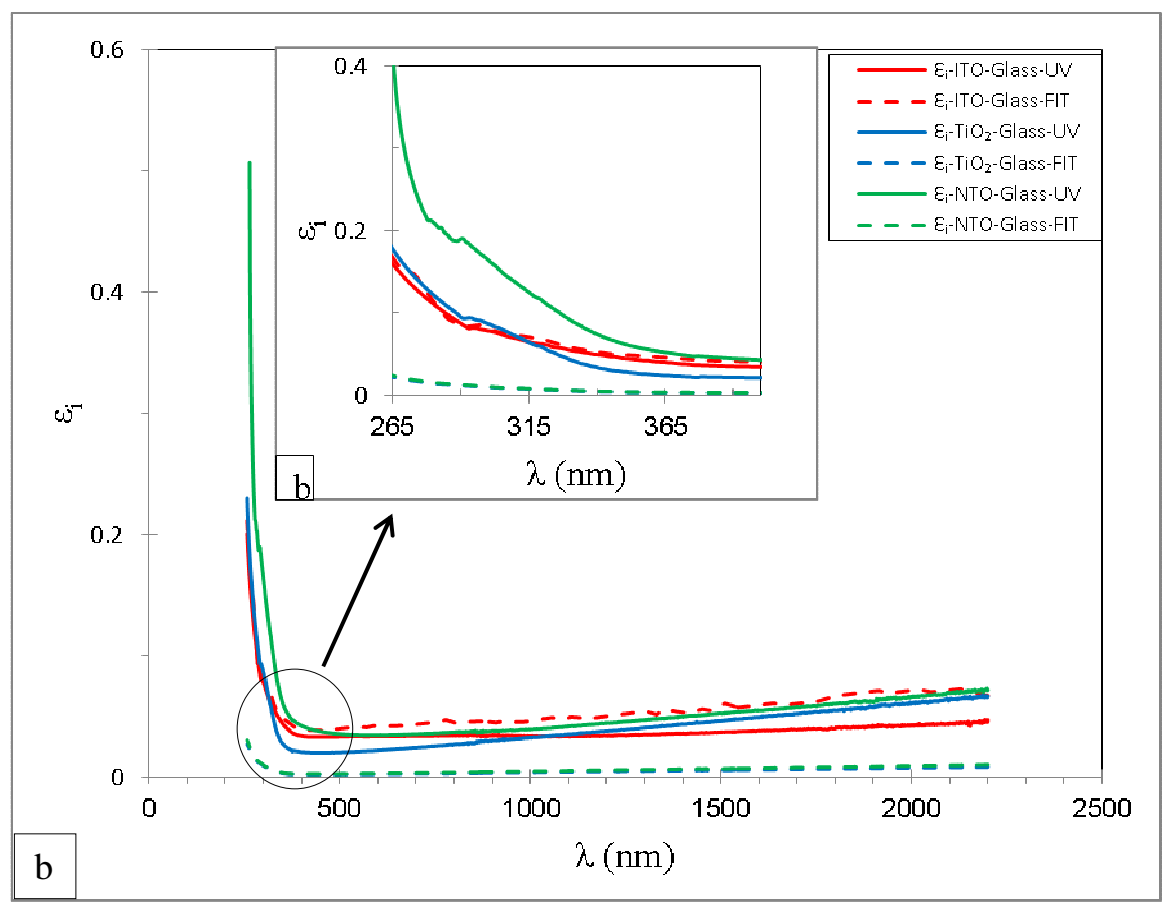

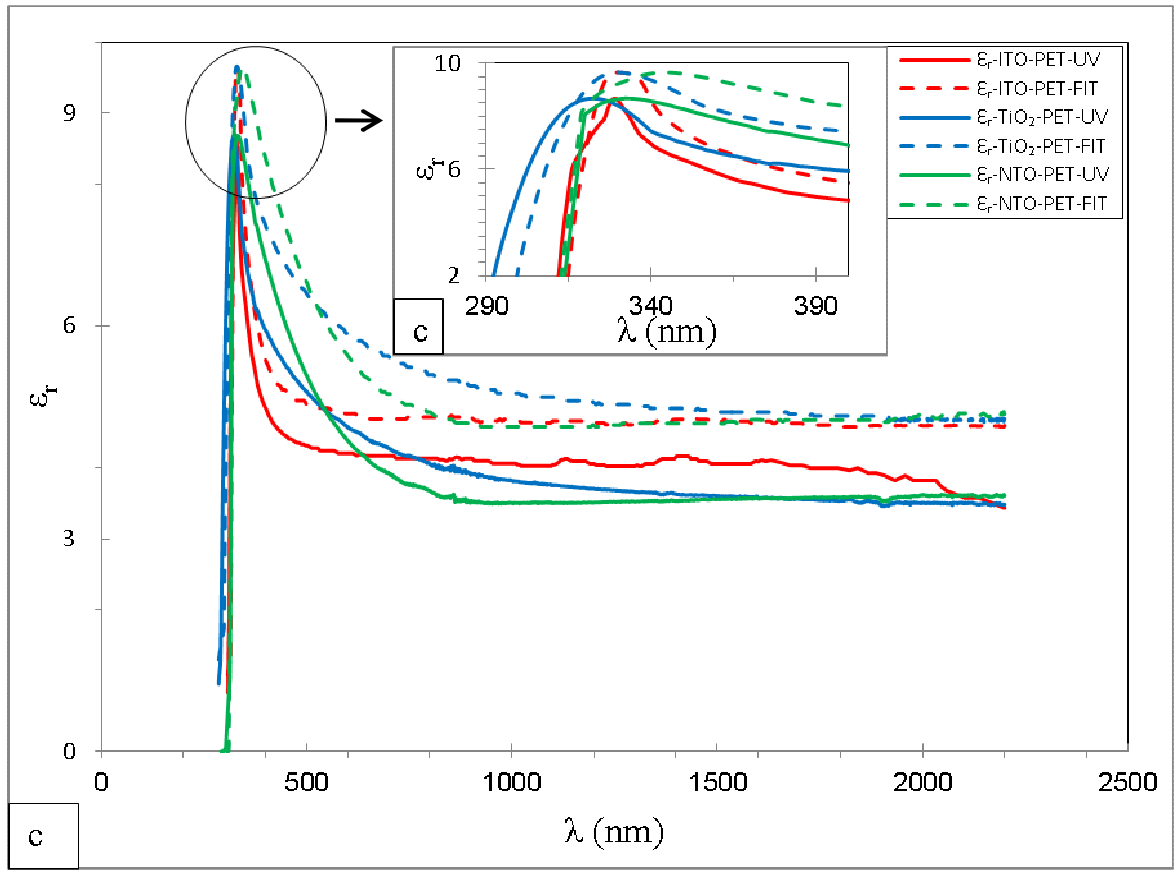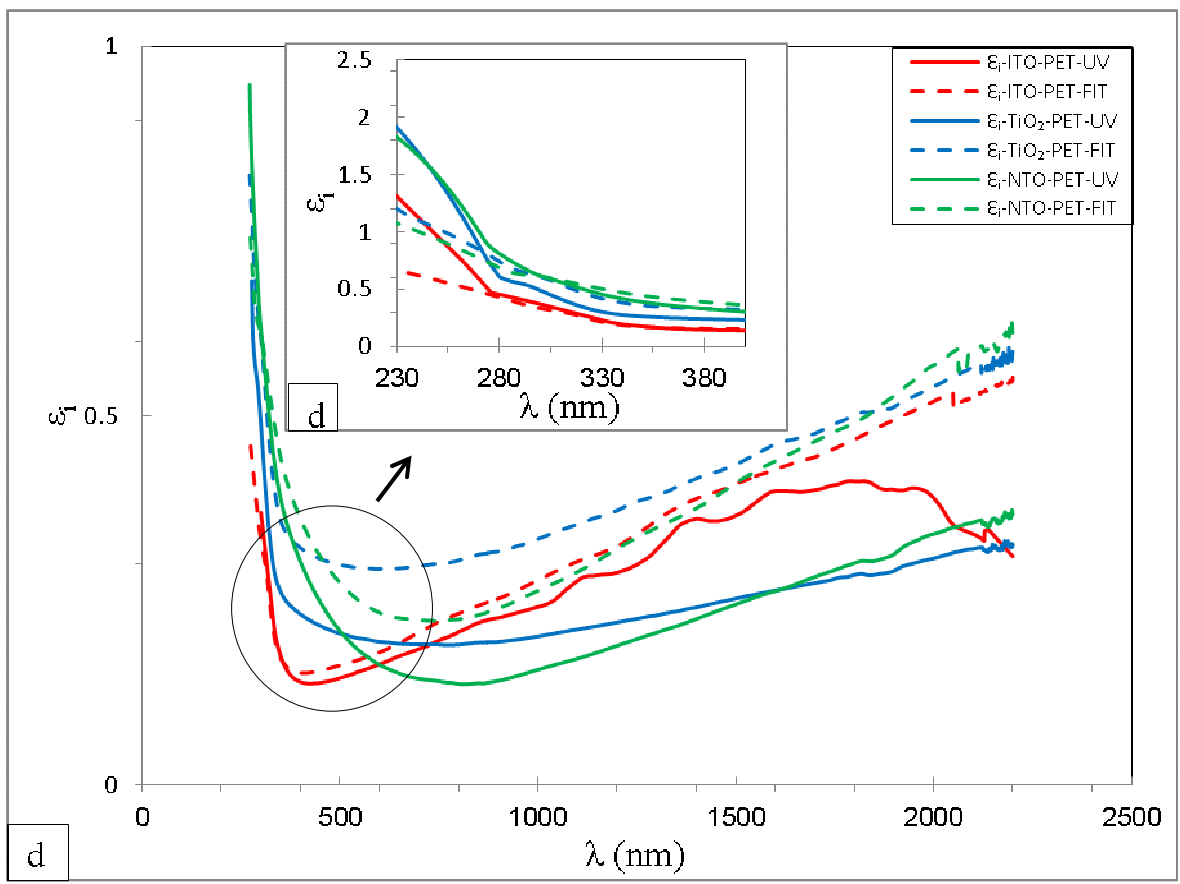

Figure 12

| Substrate | Thickness (nm) | Spectrophotometry calculations | | | | | | | |
|---|---|---|---|---|---|---|---|---|---|
| | | Glass | | | | PET | | | |
| Transition Mode | | DT | DFT | IDT | IDFT | DT | DFT | IDT | IDFT |
| ITO | 20 | 3.53 | 3.75 | 3.56 | 3.36 | 3.53 | 3.75 | 3.56 | 3.36 |
| $TiO_2$ | 20 | 3.3 | 2.9 | 2.71 | 2.2 | 3.3 | 2.9 | 2.71 | 2.2 |
| $TiO_2$:Nb | 20 | 3.6 | 3.75 | 3.03 | 3.20 | 3.6 | 3.75 | 3.03 | 3.20 |

Table 1

| Model parameters | Glass | | | PET | | |
|---|---|---|---|---|---|---|
| | ITO | $TiO_2$ | NTO | ITO | $TiO_2$ | NTO |
| $x^2$ (MSE) | 0.031 | 1.963 | 1.321 | 0.011 | 1.939 | 2.006 |
| $n_\infty$ | 1.747±0.02 | 2.154±0.04 | 13.870±2.302 | 22.123±0.982 | 2.234±0.040 | 1.859±0.02 |
| $\omega_g$ | 3.555±3.06 | 3.306±0.52 | 3.839±1.638 | 3.774±6.428 | 0.191±0.339 | 3.657±2.991 |
| $f_j$ | 0.001±0.001 | 0.0342±0.012 | 0.731±2.465 | 1.445±1.212 | 0.017±0.002 | 0.008±0.001 |
| $\omega_j$ | 9.587±0.2893 | 4.365±0.021 | 9.616±2.956 | 3.552±3.595 | 4.384±0.016 | 5.644±0.292 |
| $\Gamma_j$ | 1.381±0.134 | 0.384±0.018 | 6.463±4.755 | 4.857±2.027 | 0.369±0.015 | 2.559±1.307 |
| AOI | 72.548±0.05 | 71.664±0.042 | 70.175±1.408 | 71.733±2.027 | 71.634±0.039 | 72.565±0.048 |
| $d$ (nm) by Dektak | 20.872 | 20.365 | 20.448 | 20.896 | 20.221 | 20.777 |
| $d$ (nm) by SE | 20.113±0.945 | 20.069±0.696 | 20.764±1.953 | 20.227±1.467 | 20.917 | 20.366±0.999 |

Table 2

| Materials | Glass substrate | | | | | | | |
|---|---|---|---|---|---|---|---|---|
| | Optical constants at $\lambda_{550}$ nm | | | | | | | |
| | $n_{EXP}$ | $n_{FIT}$ | $k_{EXP}$ | $k_{FIT}$ | $\varepsilon_{r_{EXP}}$ | $\varepsilon$ | $\varepsilon$ | $\varepsilon$ |
| ITO | 1.93 | 2.06 | 0.036 | 0.042 | 3.72 | 4.3 | 0.14 | 0.17 |
| $TiO_2$ | 1.94 | 1.93 | 0.027 | 0.026 | 3.75 | 3.73 | 0.10 | 0.09 |
| NTO | 1.97 | 2.06 | 0.03 | 0.03 | 3.88 | 4.24 | 0.11 | 0.13 |

a

| Materials | PET substrate | | | | | | | |
|---|---|---|---|---|---|---|---|---|
| | Optical constants at $\lambda_{550}$ nm | | | | | | | |
| | $n_{EXP}$ | $n_{FIT}$ | $k_{EXP}$ | $k_{FIT}$ | $\varepsilon_{r_{EXP}}$ | $\varepsilon$ | $\varepsilon$ | $\varepsilon$ |
| ITO | 2.33 | 2.20 | 0.06 | 0.65 | 5.43 | 4.85 | 0.24 | 0.28 |
| $TiO_2$ | 2.37 | 2.48 | 0.06 | 0.06 | 5.61 | 6.13 | 0.27 | 0.29 |
| NTO | 2.32 | 2.45 | 0.05 | 0.06 | 5.39 | 6.02 | 0.22 | 0.16 |

b

Table 3

| Materials | Thickness (nm) | Spectroscopic ellipsometry measurements | | | | | | | |
|---|---|---|---|---|---|---|---|---|---|
| | | Glass substrate | | | | | | | |
| | | DT | | DFT | | IDT | | IDFT | |
| | | $E_{g_{EXP}}$ | $E_{g_{FIT}}$ | $E_{g_{EXP}}$ | $E_{g_{FIT}}$ | $E_{g_{EXP}}$ | $E_{g_{FIT}}$ | $E_{g_{EXP}}$ | $E_{g_{FIT}}$ |
| ITO | 21.28 | 3.4 | 3.4 | 4 | 4 | 3.4 | 3.4 | 4 | 4 |
| $TiO_2$ | 20.08 | 3.1 | 3.6 | 3.3 | 3.4 | 3.1 | 3.6 | 3.3 | 3.4 |
| NTO | 20.55 | 3.1 | 3.4 | 3.5 | 3.6 | 3.1 | 3.4 | 3.5 | 3.6 |
| Materials | Thickness (nm) | PET substrate | | | | | | | |
| | | DT | | DFT | | IDT | | IDFT | |
| | | $E_{g_{EXP}}$ | $E_{g_{FIT}}$ | $E_{g_{EXP}}$ | $E_{g_{FIT}}$ | $E_{g_{EXP}}$ | $E_{g_{FIT}}$ | $E_{g_{EXP}}$ | $E_{g_{FIT}}$ |
| ITO | 21.89 | 3.53 | 3.53 | 3.75 | 3.75 | 3.56 | 3.56 | 3.36 | 3.36 |
| $TiO_2$ | 20.09 | 3.3 | 3.3 | 2.9 | 2.9 | 2.71 | 2.71 | 2.2 | 2.2 |
| NTO | 20.23 | 3.6 | 3.6 | 3.75 | 3.75 | 3.03 | 3.03 | 3.20 | 3.20 |

Table 4